\documentclass[ilaslpaper]{ilaslarticle}
\usepackage{ilasl-en}
\usepackage{geometry}
\usepackage{amsmath}
\usepackage{amsbsy}
\usepackage{amsthm}
\usepackage{amssymb}
\usepackage{graphicx}
\usepackage{epstopdf,epsfig}
\usepackage{tikz}
\usepackage{dsfont}

\def\s{\sigma}
\def\L{\Lambda}
\def\d{\delta}

\def\righthd{Order, disorder and phase transitions in quantum many body systems}
\def\lefthd{Alessandro Giuliani}

\begin{document}

\title{Order, disorder and phase transitions in quantum many body systems}

\author{\vspace{5pt} Alessandro Giuliani}
\address{Dipartimento di Matematica e Fisica, Universit\`a degli Studi
  Roma Tre, \hfil\break L.go S.\ L.\ Murialdo 1, 00146 Roma, Italy.
  \email{giuliani@mat.uniroma3.it}
}

\date{}

\maketitle

\begin{sunto}
In queste note offro una panoramica di diversi risultati della teoria
dei sistemi quantistici a molti corpi, al confine tra la meccanica
statistica quantistica matematica e la teoria della materia
condensata. In particolare, discuto alcuni risultati recenti
sull'universalit\`a dei coefficienti di trasporto di sistemi di
elettroni interagenti, e, nello specifico, sull'indipendenza della conduttivit\`a Hall dall'interazione elettrone-elettrone. In questo contesto, lo scambio di idee tra matematica e fisica
teorica si \`e dimostrato estremamente fruttuoso e ha aiutato a chiarire
il ruolo giocato dalle leggi di conservazione quantistiche (identit\`a
di Ward) e dalle propriet\`a di decadimento delle correlazioni corrente-corrente sull'indipendenza della conduttivit\`a dall'interazione tra elettroni.
\end{sunto}

\begin{abstract}
In this paper, I give an overview of some selected results in quantum many body theory, lying at the interface 
between mathematical quantum statistical mechanics and condensed matter theory. In particular, I discuss
some recent results on the universality of transport coefficients in lattice models of interacting electrons,
with specific focus on the independence of the quantum Hall conductivity from the electron-electron interaction. 
In this context, the exchange of ideas between mathematical and theoretical physics proved particularly fruitful, and helped in clarifying 
the role played by quantum conservation laws (Ward Identities), 
together with the decay properties of the Euclidean current-current correlation functions, on the 
interaction-independence of the conductivity. 
\end{abstract}

\section{Introduction}
The attempt to understand the macroscopic properties of quantum matter starting from first principles is a central challenge in theoretical physics, which dates back to the early days of quantum 
mechanics. 
There are several exotic phenomena that are still far from being fully understood, some of which are well known since more than a century: think, for example, to superconductivity and superfluidity, 
which have first been observed in 1911 \cite{On} and 1937 \cite{AM,Ka}, respectively. Other important quantum phenomena have been discovered or observed only much more recently: think, {\it e.g.}, to the integer and fractional 
quantum Hall effect (observed in 1980 \cite{Kl} and 1982 \cite{TDG}, respectively), to high-$T_c$ superconductivity (discovered in 1986 \cite{BM}), or to Bose-Einstein condensation (observed in a gas of cold atoms in 1995 \cite{AEMWC, BSTH, DMA}). 

In several cases, approximate or heuristic theories are available for understanding the microscopic mechanism behind these exotic phenomena: I refer here, {\it e.g.}, to the BCS theory of standard 
superconductivity \cite{BCS1,BCS2}, to the Bogoliubov theory of Bose-Einstein condensation \cite{Bo}, or to the theory of the integer quantum Hall effect
\cite{La,Th, Ha, TKNdN}. Other phenomena remain less understood, 
such as high-$T_c$ superconductivity \cite{An} or the fractional quantum Hall effect \cite{La2}, whose microscopic origin is still debated, and are among the central challenges of the current research in condensed matter theory.

Even in the cases that are better understood, mathematical proofs of the quantum phase transitions of interest are typically missing, which indicates an incomplete understanding of the subject: 
for instance, in standard superconductivity, there is no systematic way of controlling the corrections beyond mean field, which is the basic approximation that BCS theory is based on. 
In the context of Bose-Einstein condensates, a full proof of condensation for a homogenous gas in the thermodynamic limit is missing, which is a signal of our poor understanding of 
the phenomenon of continuous symmetry breaking in quantum many-body systems. In the context of the quantum Hall effect, a proof of the stability of the Hall plateaux in the presence of 
disorder and electron-electron interactions has still to come, which is a signal of our poor understanding of the interplay between disorder and many-body interactions. 

Notwithstanding these limitations, the last years witnessed very important developments in the rigorous understanding of these elusive phenomena, at the interface between mathematical and theoretical 
physics. By combining the ideas developed in the last decades in the condensed matter community, including the use of Ward Identities in formal perturbation theory, the use of 
effective field theories, and the proposal of 
geometrical indices characterizing the `quantum topological phases', 
with sophisticated mathematical tools, such as functional inequalities, localization bounds, 
constructive field theory and multiscale analysis, we acquired a better understanding of several remarkable phenomena, including, {\it e.g.}, Bose-Einstein condensation \cite{LSSY}, quantum magnetism \cite{CGS}, and the universality of quantum transport coefficients \cite{GJMP,GMP11,GMP17,HM}, just to mention a few. 

In this paper I will review some of the latest developments in the universality theory of quantum transport coefficients, which allowed to clarify certain debated issues connected with the optical conductivity in graphene \cite{GMP11}. The key new technical tool that emerged from the combination of theoretical and mathematical physics ideas, is the implementation of Ward Identities 
within the constructive scheme (`multiscale fermionic cluster expansion') that is currently able to control the analyticity and decay of correlation functions for the ground state of several two-dimensional 
interacting electron systems. Note that a formal use of Ward Identities in the effective field theory description of quantum phenomena can easily lead to inconsistent results, particularly as far as 
the computation of transport coefficients is concerned, cf., {\it e.g.}, with \cite{HJV}. 

In order to make the ideas behind these recent applications as transparent as possible, I will restrict my attention to the study of the universality properties of the Kubo conductivity, and, in particular, of its 
transverse component (Hall conducitivity) in weakly interacting lattice fermions characterized by a gapped (`massive') reference non-interacting Hamiltonian. 
For these systems, the construction of the ground state correlation functions and the 
proof of their analyticity properties is particularly simple and, strictly speaking, does not require a multiscale expansion at all. The extension of the proof to the gapless case, in particular in the case
 of graphene-like systems, requires the use of a multiscale analysis (constructive fermionic renormalization group), which goes beyond the purpose of this review. 
 
The argument presented here is based on \cite{GMP17}, which I will refer to for some technical aspects of the proof. However, compared to \cite{GMP17}, the proof presented here has some
important simplifications in the proof of analyticity and exponential decay of correlations, as well as in the use of the Schwinger-Dyson equation. 

The plan is to first introduce the context, the model and the main results. Next, I will present the proof, first giving an overview of the structure of the proof, and then 
explaining in some details the different steps. 

\section{The quantum Hall effect}

Before presenting the main results, let me clarify the context under consideration: the quantum Hall effect is a peculiar electronic transport phenomenon, which is observed in thin
conducting, or semi-conducting, materials. By `thin', here, I mean that the material samples under consideration are two-di\-men\-sional, or quasi-two-dimensional. The quantum phenomenon 
of interest has a classical counterpart, which is important to keep in mind: the classical Hall effect is observed in thin conducting materials subject both to a longitudinal electric field $E$ (along the direction of the 
current) and to a transverse magnetic field $B$, as shown in the picture. 

\begin{figure}[ht]
\centering
\includegraphics[width=\textwidth]{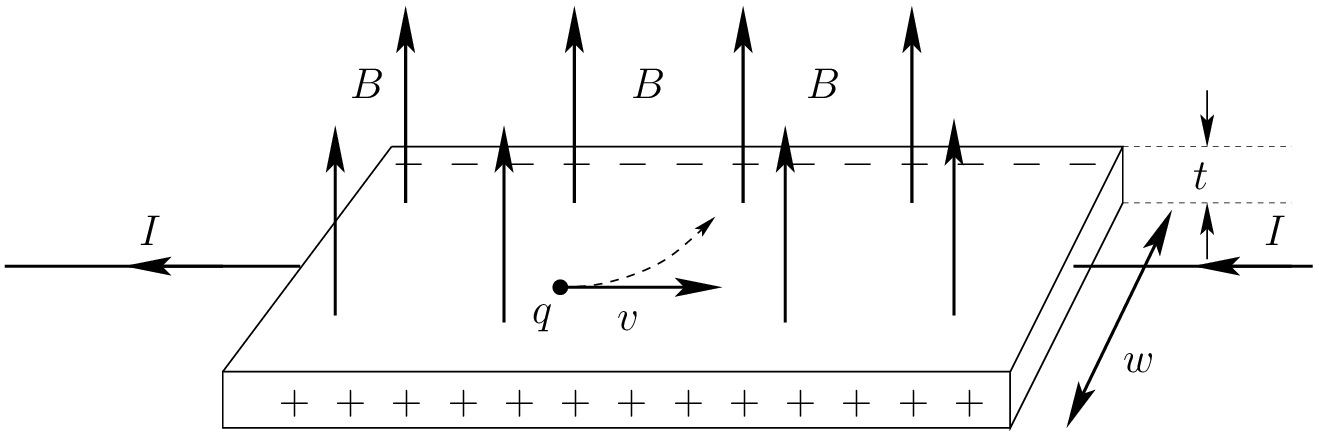}
\caption{\it A sketch of the setting for the Hall effect: a steady longitudinal current $I$ flows in a thin conducting material, of 
thickness $t$ and width $w$, in the presence of a transverse magnetic field $B$. In the picture, the charge $q$ is negative, and the deflection 
due to the Lorenz force is schematically represented.}\label{fig.1}
\end{figure}

Let $t$ be the film thickness, and $w$ the sample width. As the electrons move in the direction of the current, due to the presence of a transverse magnetic field, they are subject to the Lorentz force
$F_L=qv\wedge B$ (here $q$ is the electron charge and $v$ its velocity), which tends to deflect them towards the edges of the sample. In this way, electrons start to accumulate at the edges, and the accumulation goes on until the transverse electric field produced by these electrons is strong enough to compensate the Lorentz force exactly. The equilibrium transverse voltage that is generated in this way is known as the Hall voltage \cite{Hall1879} and is equal to:
$$V_H=vBw.$$
By rewriting the electron velocity as $v=I/(n^{3d}qtw)$, where $I$ is the intensity of the current, and $n^{3d}$ the (three-dimensional) charge carriers density, one gets
$$V_H=\frac{IB}{n^{3d}qt}.$$
Note that $V_H$ is inversely proportional to the film thickness, which explains why the Hall effect is easier to observe in thin samples than in thick ones. For two-dimensional samples, we get the analogous formula: $V_H=IB/(nq)$, where $n=n^{2d}$ is the two-dimensional charge carriers density. Letting $I=jw$, where $j$ is the sheet current density, and defining the Hall field as $E_H=V_H/w$, one finds that the Hall conducitivity has the following expressions: 
$$\s_H=\frac{j}{E_H}=\frac{nq}{B}.$$
Often, this formula for $\s_H$ is equivalently rewritten in `natural units' as
$$\s_H=\frac{q^2}{h}\nu,$$
where $h$ is Planck's constant, and, if $\Phi_0=h/q$ is the flux quantum, $\nu=n\Phi_0/B$ is called the `filling factor'. Written in these terms, the Hall law is a linear relation between the 
transverse conductivity and the filling factor, with an explicit proportionality coefficient, equal to the `conductivity quantum' $q^2/h$.

Experimentally \cite{Kl} it turns out that at low temperatures and/or large magnetic fields, when the filling factor is of the order of a few units, the system displays a quantization effect: $\s_H$ displays 
`plateaus' on which $\s_H$ is constant at a remarkable precision ($\sim 10^{-9}$). This quantization effect is well understood, at a mathematically rigorous level, in the case of non-interacting electron models, 
including, possibly, disorder, {\it i.e.}, a random potential \cite{AG, ASS, BVS,TKNdN}. 

The key observation behind the mathematical theory of the integer quantum Hall effect is the interpretation of the transverse Kubo conductivity as a geometric index (first Chern number of the Bloch bundle). 
Unfortunately, the representation breaks down in the presence of electron-electron interactions. 

In this review I will give an introduction to the theory of the {\it interacting} quantum Hall effect, for a class of two-dimensional lattice electron models with weak interactions and no disorder. For more general 
models, in particular for larger values of the interaction strength, new quantization effects are expected to take place (fractional quantum Hall effect \cite{STG}).

\section{The models}

The models that I will consider are defined as follows. Consider a finite portion $\L$ of a two-dimensional Bravais lattice generated by two independent vectors $\ell_1,\ell_2\in\mathbb R^2$. 
I will assume for definiteness that periodic conditions are imposed at the boundary, and that the system is periodic of period $L$ in both directions $\ell_1$ and $\ell_2$, so that 
$|\L|=L^2$. At each site $x$ of the lattice I associate a finite number of fermionic creation/annihilation operators, $\psi^\pm_{x,\s}$, with `color' $\s$ taking values in a finite index set $I$. In typical examples, $\s$ can represent a spin index, and/or the position index within the unit cell associated with $x$. 

The `configuration space' consists of vectors in a fermionic Fock space, labelled by the occupation numbers ${\underline n}=\{n_{x,\s}\}_{x\in\L, \s\in I}$ of the fermions, with $n_{x,\s}=0,1$. 
The canonical anti-commutation rules for the creation/ annihilation operators are $\{\psi^+_{x,\s},\psi^+_{y,\s'}\}= \{\psi^-_{x,\s},\psi^-_{y,\s'}\}=0$, and 
$$\{\psi^+_{x,\s},\psi^-_{y,\s'}\}=\d_{x,y}\d_{\s,\s'}.$$
The vector in Fock space representing the state with occupation number $\underline n$ is obtained by acting on the fermionic vacuum $|0\rangle$ with $\prod_{x,\s}(\psi^+_{x,\s})^{n_{x,\s}}$,
where the product of the different, mutually anti-commuting, operators, must be performed in some prescribed order, fixed once and for all.

The grand-canonical Hamiltonian of the system will be assumed to have the following form: 
\begin{equation}H=H_0+UV-\mu N,\label{Ham}\end{equation}
where 
$$H_0=\sum_{x,y\in\L}\sum_{\s,\s'\in I}\psi^+_{x,\s}H^0_{\s,\s'}(x-y)\psi^-_{y,\s'},$$
$$V=\sum_{x,y\in\L}\sum_{\s,\s'\in I}n_{x,\s}v_{\s,\s'}(x-y)n_{y,\s'},\qquad {\rm with}\qquad n_{x,\s}=\psi^+_{x,\s}\psi^-_{x,\s}, $$
and
$$N=\sum_{x\in \Lambda}\sum_{\sigma\in I}n_{x,\s},$$
In these formulas, $H_0$ is the `hopping Hamiltonian', and $H^0_{\s\s'}(x-y)$ are the `hopping strengths', assumed to be of finite range and such that $H^0_{\s\s'}(x)=[H^0_{\s'\s}(-x)]^*$.
The operator $V$ has the interpretation of a density-density interaction, with the coefficients $v_{\s\s'}(x)$ that are assumed to be real, symmetric ({\it i.e.}, $v_{\s\s'}(x)=v_{\s'\s}(-x)$), and of finite range.
Finally, $\mu$ is the chemical potential, to be fixed in a spectral gap of the unperturbed Hamiltonian, as discussed in the following. 

\looseness +1
The spectral properties of $H_0$ are most conveniently described in terms of the `Bloch Hamiltonian' $\hat H^0(k)$, which is the Fourier transform of the $|I|\times|I|$ hopping matrix $H^0(x)$: 
$$\hat H^0(k)=\sum_{x\in\L}e^{-ikx}H^0(x),$$
where $k$ is an element of the discretized Brillouin zone, 
$$\mathcal{B}_{L} := \Big\{ k\in\mathbb R^2\ : \ k = \frac{n_{1}}{L} G_{1} + \frac{n_{2}}{L} G_{2},\; n_{i}\in [0,L)\cap\mathbb{Z} \Big\}\,,$$
and $G_1,G_2$ form a basis of the dual lattice $\L^*$, {\it i.e.}, $G_i\cdot\ell_j=2\pi\delta_{ij}$. I will also denote $\mathcal B\equiv\mathcal B_\infty$. 

If, as assumed, $H^0(x)$ has finite range, then the (thermodynamic limit of the) Bloch Hamiltonian is a Hermitian matrix, depending {\it analytically} on the momentum $k$ over the 
Brillouin zone $\mathcal B$, with periodic conditions at the boundary
of $\mathcal B$. The  Bloch Hamiltonian, $\sigma(\hat
H^0(k)) =\{\epsilon_\s(k)\}_{\s\in I}$ has a real spectrum, and the 
functions $\epsilon_\s(k)$ are known as the `energy bands'. 

I assume the chemical potential to be in a spectral gap: 
\begin{equation}\label{gapcon}\delta_\mu=\inf_{k\in\mathcal B}{\rm dist}\big(\mu,\sigma(\hat H^0(k))\big)>0.\end{equation}
An important fact is that, if $\delta_\mu>0$, then the projector $P_-(k)$ over the filled bands, defined as
$$P_-(k)=\sum_{\alpha\ :\ \epsilon_\alpha(k)<\mu}P_\alpha(k),$$
with $P_\alpha(k)$ the projector over the $\alpha$-th band, is {\it analytic} in $k$. 

\medskip

\noindent
{\bf Observables\quad} $O$ correspond to self-adjoint operators on the fermionic Fock space, and their expectation value at inverse temperature $\beta$ is
$$\langle O\rangle_{\beta,L}=\frac{{\rm Tr}\ e^{-\beta H}O}{{\rm Tr}\ e^{-\beta H}}$$
I denote by $\langle\cdot\rangle$ the thermodynamic and zero temperature limit of $\langle\cdot\rangle_{\beta,L}$: 
$$\langle O\rangle=\lim_{\beta\to\infty}\lim_{L\to\infty}\langle O\rangle_{\beta,L}.$$
The goal is to characterize the properties of the interacting, infinite-volume, ground state $\langle\cdot\rangle$, in terms of its `correlation functions' 
$\langle O_1\cdots O_n\rangle$, with $O_i$ being fermionic operators depending on a finite number of fermionic fields. I will be particularly interested in current-current correlations, 
in terms of which one can compute the {\it Kubo conductivity}, to be defined in the following. 

Before that, let me discuss a couple of explicit examples of hopping Hamiltonian $H_0$, which are characterized by a non-trivial behavior of the Hall conductivity, as the free parameters of the model are properly varied. 

\subsection{Examples}
{\it The Hofstadter model with rational flux.} The non-interacting version of this model describes tight-binding electrons, hopping between nearest neighbor sites of a square lattice, in the presence of an external, constant, 
transverse magnetic field. I let $\Lambda$ be a square lattice, with $\ell_1=(1,0)$ and $\ell_2=(0,1)$, and  
$$H_0=-t\sum_{x\in\Lambda}\sum_{i=1,2}\psi^+_x\psi^-_{x+\ell_i}e^{iq\int_x^{x+\ell_i}A\cdot d\ell}+H.c.,$$
where $H.c.$ indicates the Hermitian conjugate, $t>0$ is the hopping strength, and $A$ is a vector potential corresponding to a constant transverse magnetic field. Choosing, {\it e.g.}, $A=(0,Bx_1)$, the hopping Hamiltonian 
takes the form: 
$$H_0=-t\sum_{x\in\Lambda}\Big(\psi^+_x\psi^-_{x+\ell_1}+\psi^+_x\psi^-_{x+\ell_2}e^{iqBx_1}\Big)+H.c.$$
I assume $qB$ to be a rational multiple of $2\pi$, {\it i.e.}, $qB=2\pi m/n$, with $m,n$ relative prime integers, and the system size $L$ to be proportional to $n$.
Note that the hopping coefficients are invariant under translations by multiples of $(n,1)$. Therefore, it is convenient to introduce a unit cell labelled by $n$ colors, $\sigma\in\{1,\ldots,n\}$,
so that $H_0$ can be re-written, after a natural re-labeling of the field and site indices, as
$$H_0=-t\sum_{x\in\Lambda}\Big[\sum_{\sigma=1}^{n-1}\psi^+_{x,\sigma}\psi^-_{x,\sigma+1}+\psi^+_{x,n}\psi^-_{x+\ell_1,1}+\sum_{\sigma=1}^n\psi^+_{x,\sigma}\psi^-_{x+\ell_2,\sigma}
e^{i2\pi m\sigma/n}\Big].$$
The structure of the energy bands depends in a very peculiar (fractal) way on the value of $m/n$: more precisely, the value of the transverse Kubo conductivity (to be defined below),
if plotted against the magnetic field $qB=2\pi m/n$ and the chemical potential $\mu$ displays a fractal structure, which gives rise to the famous {\it Hofstadter butterfly} \cite{Ho}. The band theory of this model is very 
rich and interesting, see, {\it e.g.}, \cite{Av} for details. 

\bigskip

{\it The Haldane model.} The non-interacting version of this model describes tight binding electrons, hopping between nearest-neighbor and next-to-nearest-neighbor sites of an hexagonal lattice,
in the presence of an external, {\it dipolar}, transverse magnetic field, as well as of a staggered chemical potential. 
I think the hexagonal lattice as the union of two translated copies of a triangular lattice $\Lambda$, with $\ell_{1,2}=(3/2,\mp\sqrt3/2)$.
The sites of the two copies can be represented as being white (corresponding to a `color label' $\sigma=1$) and black (corresponding to a `color label' $\sigma=2$).
I think $\Lambda$ as coinciding with the sub-lattice of white sites, and the black sites as belonging to $\Lambda+(1,0)$. 
The creation/annihilation operators associated with the white site located at $x\in\Lambda$ will be denoted by $\psi^\pm_{x,1}$,
while the same operators associated with the black site located at $x+(1,0)$ will be denoted by $\psi^\pm_{x,2}$.

The hopping Hamiltonian of the Haldane model is defined as follows: 
\begin{eqnarray}
H_0&=&-t_1\sum_{x\in\Lambda}\Big[\psi^+_{x,1}\big(\psi^-_{x,2}+\psi^-_{x-\ell_1,2}+\psi^-_{x-\ell_2,2}\big)+H.c.\Big]\nonumber\\
&&-t_2\sum_{x\in\Lambda}\sum_{j=1}^3\sum_{\alpha=\pm}\Big(e^{i\alpha\phi}\psi^+_{x,1}\psi^-_{x+\alpha\gamma_j,1}+
e^{-i\alpha\phi}\psi^+_{x,2}\psi^-_{x+\alpha\gamma_j,2}\Big)\nonumber\\
&&+M\sum_{x\in\Lambda}\Big(\psi^+_{x,1}\psi^-_{x,1}-\psi^+_{x,2}\psi^-_{x,2}\Big),\nonumber\end{eqnarray}
where: $\gamma_1=\ell_1-\ell_2$, $\gamma_2=\ell_2$, $\gamma_3=-\ell_1$; $t_1>0$ and $t_2>0$ are the nearest-neighbor and next-to-nearest-neighbor hopping strengths; 
$\phi$ is a parameter that measures the strength of the dipolar magnetic field; and $M$ is the amplitude of the staggered chemical potential. 

\begin{figure}[ht]
\centering
\includegraphics[width=0.8\textwidth]{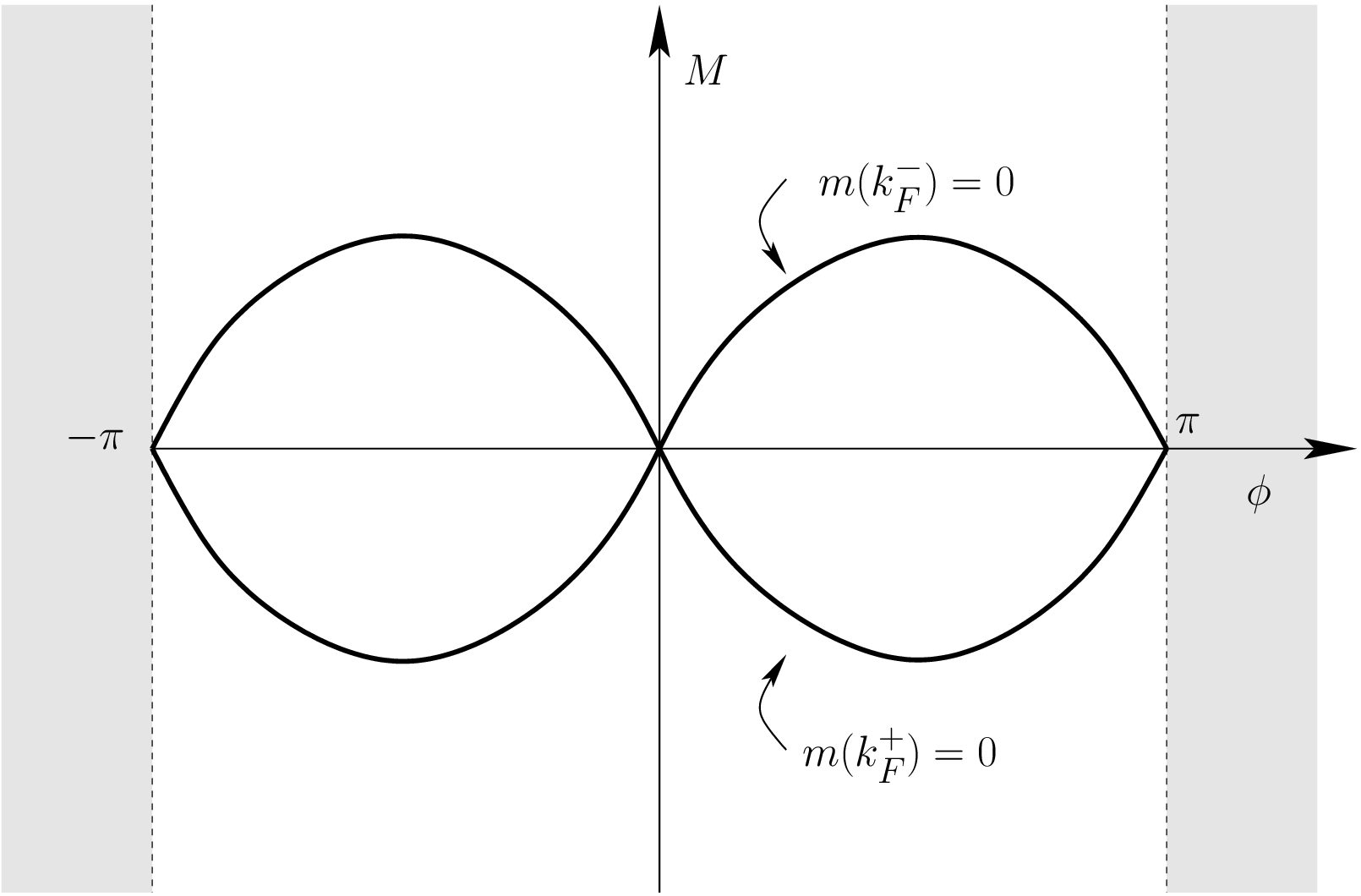}
\caption{\it  The critical lines $m(k_F^\pm)=0$ of the Haldane model in the $(\phi,M)$ plane.}
\label{fig.2}
\end{figure}

The nice feature of this model, which makes it a very interesting playground, is that the energy bands are very easy to calculate, and so is the Kubo conductivity. Moreover, as we shall see, 
the transverse conductivity has a non-trivial behavior, as the parameters $M$ and $\phi$ are varied. In fact, the Bloch Hamiltonian takes the form: 
$$\hat H^0(k)=\begin{pmatrix} -2t_2\alpha_1(k)\cos\phi+m(k)& -t_1\Omega^*(k)\\
-t_1\Omega(k) & -2t_2\alpha_1(k)\cos\phi-m(k)\end{pmatrix},$$
where $\alpha_1(k)=\sum_{j=1}^3\cos k\cdot \gamma_j$, and $m(k)=M-2t_2\alpha_2(k)\sin\phi$, with $\alpha_2(k)=\sum_{j=1}^3\sin k\cdot \gamma_j$. Moreover, $\Omega(k)=1+e^{-ik\cdot\ell_1}+
e^{-ik\cdot\ell_2}$. Note that $|\Omega(k)|$ vanishes if and only if $k=(\frac{2\pi}3,\pm\frac{2\pi}{3\sqrt3})\equiv k_F^\pm$: the two points $k_F^\pm$ are called the {\it Fermi points}. 
The energy bands associated with the Bloch Hamiltonian are: 
$$\epsilon_\pm(k)=-2t_2\alpha_1(k)\cos\phi\pm\sqrt{t_1^2|\Omega(k)|^2+m^2(k)}.$$
I assume that $t_2/t_1<1/3$: in this way, using the fact that $\max_k|\Omega(k)|=|\Omega(0)|=3$ and that $\alpha_1(k)=|\Omega(k)|^2/2-3/2$, one sees that the two bands can touch only if
$m(k_F^\pm)=M\pm 3\sqrt3 t_2\sin\phi=0$, which is the equation for the
{\it critical lines} of the Haldane model. The graph of the critical,
massless, lines in the $(\phi,M)$ plane is shown in {\it Fig. \ref{fig.2}}

For values of $(\phi,M)$ in the complement of the two critical lines, the two bands $\epsilon_\pm(k)$ are separated by a gap, so that $\mu$ can be chosen in such a way that the gap condition \eqref{gapcon} is satisfied. 
Note that the complement of the critical lines ({\it i.e.}, the gapped region) is naturally partitioned in four disconnected portions, corresponding to the cases $m(k_F^+)>0, m(k_F^-)>0$, or 
$m(k_F^+)>0>m(k_F^-)$, or $m(k_F^+)<0<m(k_F^-)$, or $m(k_F^+)<0, m(k_F^-)<0$.

\section{Current and conductivity}

The total (d.c.) current of an electron system is defined as $J=q \hat v$, with $\hat v=i[H, X]$ the velocity of the system: here $ X$ is the position operator, which takes the form
$$X=\sum_{x\in\Lambda}\sum_{\sigma\in I}n_{x,\sigma}(x+r_\sigma)$$
and $r_\sigma$ represents the position of the site of color $\sigma$, relative to the position $x$ of the cell it belongs to. Using the definitions of $\hat v,H$ and $X$, one obtains: 
$$J=-iq\sum_{{x,y\in\Lambda}}\sum_{
\sigma,\sigma'\in I}\psi^+_{x,\sigma} H^0_{\sigma,\sigma'}(x-y)\psi^-_{y,\sigma'}(x-y+r_\sigma-r_{\sigma'}).$$
The {\bf Kubo conductivity} is the linear response coefficient of the current to a small external electric field $E$, which is assumed to be uniform in space and
adiabatically switched on in the far past. More precisely, I introduce the time-dependent Hamiltonian $H(t)$, which is obtained from $H$ as follows: 
$$H(t)=H-qe^{\eta t}\sum_{\substack{x\in\Lambda \\\sigma\in I}}n_{x,\sigma}E\cdot (x+r_\sigma), \qquad t\in(-\infty,0].$$
Here $\eta>0$ is an adiabatic parameter, to be eventually sent to zero. I assume that, at $t=-\infty$, the density matrix of the system is the equilibrium one, associated with the Hamiltonian $H$ and the inverse temperature $\beta$: $$\rho(t)\big|_{t=-\infty}=e^{-\beta H}\equiv \rho_{-\infty}.$$ For any $t\le 0$, the density matrix $\rho(t)$ is defined as the solution to the differential equation
$$\dot \rho(t)=-i[H(t),\rho(t)],$$
which gives, at first order in $E$,
$$\rho(t)=\rho_{-\infty}+iq\int_{-\infty}^t e^{\eta \tau}e^{-iH(t-\tau)}[E\cdot X,\rho_0]e^{iH(t-\tau)}d\tau+O(E^2).$$
Therefore, the mean value of the current per unit area at time $t=0$, at first order in $E$, in the thermodynamic, zero temperature and adiabatic limits, is: 
$$\lim_{\eta\to 0^+}\lim_{\beta\to\infty}\lim_{L\to\infty}\frac{iq}{|\ell_1\wedge\ell_2|}\frac1{L^2}
\int_{-\infty}^0 e^{\eta \tau}
\frac{{\rm Tr} J e^{iH\tau}[E\cdot X,e^{-\beta H}]e^{-i H\tau}}{{\rm Tr}e^{-\beta H}}d\tau.$$
Correspondingly, the linear response coefficient of the $i$-th component of the current $J_i$ to the $j$-th component of the 
electric field $E_j$ is 
$$\sigma_{ij}=-\frac{q^2}{|\ell_1\wedge\ell_2|}\int_{-\infty}^0 e^{\eta \tau}\pmb{\langle} \Big[e^{-iH\tau}[H,X_i]e^{iH\tau},X_j\Big]\pmb{\rangle}_{\infty}d\tau,$$
where $\pmb{\langle}\cdot\pmb{\rangle}_{\infty}=\lim_{\beta\to\infty}\lim_{L\to\infty}L^{-2}\langle\cdot\rangle_{\beta,L}$. The matrix of elements $\sigma_{ij}$ is known as the Kubo 
conductivity matrix. It is customary to rewrite $\sigma_{ij}$ in a different, equivalent, form. By using the Leibniz rule for the commutator, one finds
$$
\displaylines{\quad
  \pmb{\langle}  \Big[e^{-iH\tau}[H,X_i]e^{iH\tau},X_j\Big]\pmb{\rangle}_{\infty}
  = \pmb{\langle} [e^{-iH\tau},X_j]\, [H,X_i]
  e^{iH\tau}\pmb{\rangle}_{\infty}
\hfill\cr\hfill
  +\pmb{\langle} \Big[[H,X_i],X_j\Big]\pmb{\rangle}_{\infty}\nonumber
+\pmb{\langle} e^{-iH\tau}[H,X_i]\, [e^{iH\tau},X_j]\,
\pmb{\rangle}_{\infty}.\quad}
$$
Moreover, $ [e^{-iH\tau},X_j]$ can be rewritten as
$$ [e^{-iH\tau},X_j]=-i\int_0^\tau e^{-iHt}[H,X_j]e^{iH(t-\tau)}dt.$$
By using these relations in the expression of $\sigma_{ij}$, and recalling the definition of $J_i$,  one gets: 
\begin{eqnarray}
\sigma_{ij}&=& \frac1{|\ell_1\wedge\ell_2|}
\lim_{\eta\to0^+}\frac1{\eta}\,\times\label{eq.0}\\
&\times&\Bigl\{
-i\int_{-\infty}^0e^{\eta t}\pmb{\langle} [J_i,J_j(t)]\pmb{\rangle}_{\infty}-q^2\pmb{\langle}\Big[[H,X_i],X_j\Big]\pmb{\rangle}_\infty\Bigr\},
\nonumber\end{eqnarray}
where $J_j(t)=e^{iHt}J_je^{-iHt}$. This is one of the standard expressions for the Kubo conductivity. The second term in braces is known as the diamagnetic, or Schwinger, term. 

Note that $\sigma_{ij}$ is a function of the parameters entering the Hamiltonian $H$ 
of the system, in particular of the strength $U$ of the electron-electron interaction. 
The main result reviewed in these notes is the following.

\bigskip

\noindent
{\bf Theorem 1 \cite{GMP17}.} {\it In the context above, let $\delta_\mu>0$. There exists $U_0>0$ such that, if $|U|\le U_0$, then 
$$\sigma_{ij}=\sigma_{ij}\Big|_{U=0},$$
that is, if the chemical potential is chosen so that the gap condition is verified, then the Kubo conductivity is independent of the strength of the interaction $U$.}

\bigskip

\noindent
    {\bf Remarks.}

\medskip\noindent
1) Under the gap condition, the non-interacting Kubo conductivity can be re-written as $q^2$ times the first Chern number of the Bloch bundle, {\it i.e.}, of the vector bundle associated with the linear space 
Ran$P_-(k)$, with $k\in\mathcal B$: 
\begin{equation}\label{eq.2}\sigma_{ij}\Big|_{U=0}=-iq^2\int_{\mathcal B}{\rm Tr}\, P_-(k)[\partial_{k_i}P_-(k),\partial_{k_j}P_-(k)] \frac{dk}{(2\pi)^2},\end{equation}
which is known to be proportional to an integer \cite{ASS,TKNdN}. 
More precisely, $\sigma_{11}\big|_{U=0}=\sigma_{22}\big|_{U=0}=0$, while $\sigma_{12}\big|_{U=0}=-\sigma_{21}\big|_{U=0}\in\frac{q^2}{2\pi}\mathbb Z$. 
The off-diagonal Kubo coefficient $\sigma_{12}\big|_{U=0}$ is the non-interacting {\it Hall conductivity}, and the representation in terms of a Chern number shows that it is quantized. 
Theorem 1 shows that, under the gap condition and at weak enough coupling, the interacting Hall conductivity is also quantized, at the very same value as the reference non-interacting system. 
A sketch of the proof of the formula \eqref{eq.2} is given in Appendix \ref{app.A}.

\medskip

\begin{figure}[ht]
\centering
\includegraphics[width=0.8\textwidth]{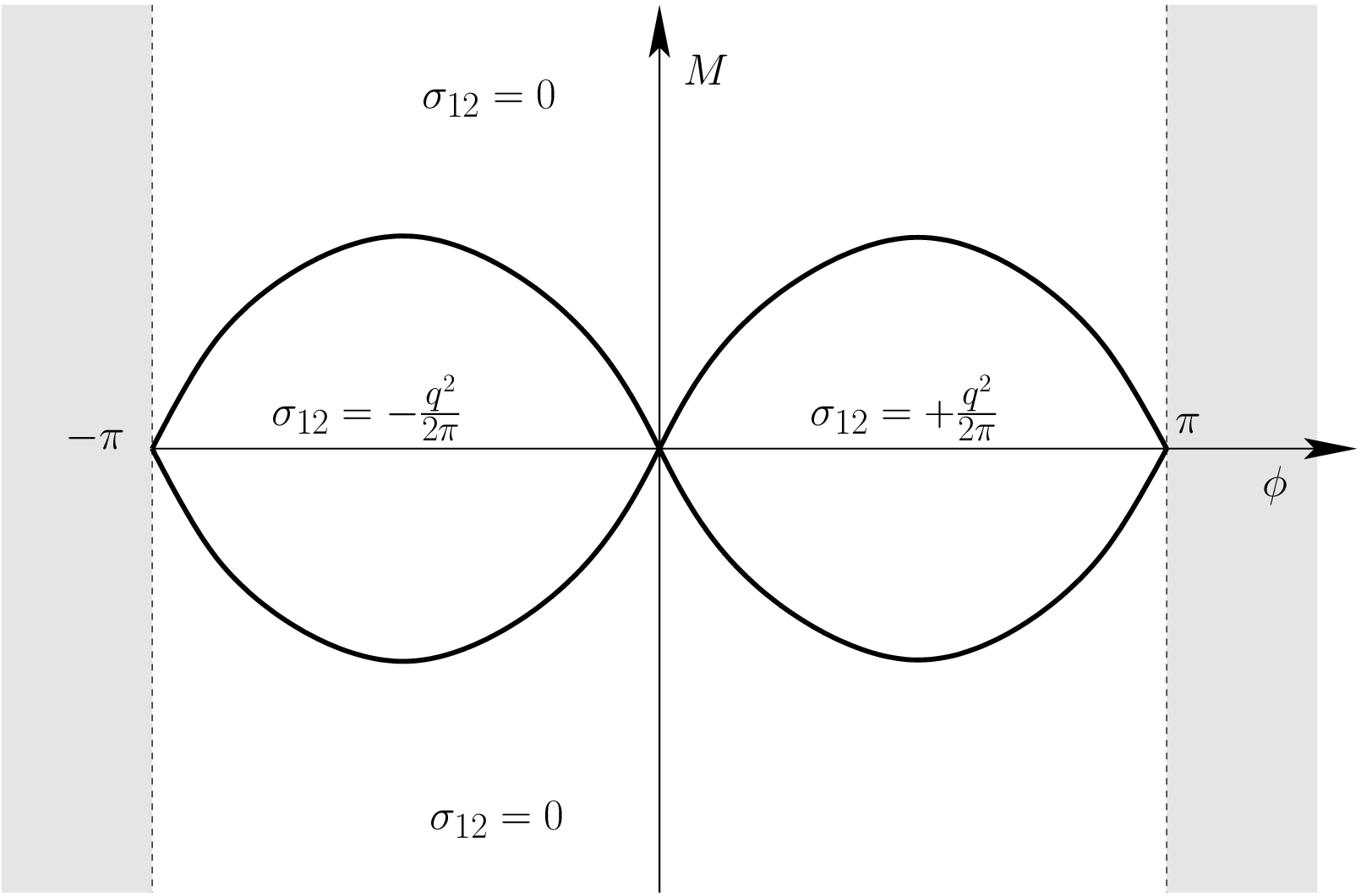}
\caption{\it The `topological phase diagram' of the Haldane model, where the different values of $\sigma_{12}$ in the four regions 
$\{m(k_F^+)>0, m(k_F^-)>0\}$,  
$\{m(k_F^+)>0>m(k_F^-)\}$, $\{m(k_F^+)<0<m(k_F^-)\}$, and $\{m(k_F^+)<0, m(k_F^-)<0\}$ are indicated.}
\label{fig.3}
\end{figure}

\noindent2) In order to prove that the Hall conductivity is non-trivial ({\it i.e.}, different from zero), one needs to perform an explicit computation that, of course, is model-dependent. 
The two examples mentioned above, {\it i.e.}, the Hofstadter and the Haldane models, are among the simplest examples of gapped systems displaying a non-trivial Hall 
conductivity. The computation of the value of the Hall conductivity for the {\bf Hofstadter model} as a function of the magnetic flux is a very non-trivial and interesting exercise: the computation is 
reduced to the study of a Diophantine equation, which can be solved numerically. The resulting `topological phase diagram' ({\it i.e.}, the plot of the value of the Hall conductivity as the magnetic flux and the
chemical potential are varied) leads to the well known `Hofstadter butterfly', whose construction goes beyond the purpose of this review \cite{Av,Ho,TKNdN}. The computation of the Hall conductivity for the {\bf Haldane model} is much simpler: in fact, the evaluation of $\sigma_{12}\big|_{U=0}$ based on \eqref{eq.2} is recommended as a very instructive exercise. The interested reader can consult, {\it e.g.}, 
\cite[Appendix B]{GMP17}. The resulting topological phase diagram is
shown in {\it Fig. \ref{fig.3}}. It first appeared in \cite{Hal}.

\medskip

\noindent3) The range of interaction strengths for which Theorem 1 holds, $|U|\le U_0$, is, in general, non uniform in the spectral gap $\delta_\mu$. 
However, in specific models, it is possible to obtain estimates uniform in the gap: 
{\it e.g.}, for the interacting Haldane model, it was proved in \cite{GJMP,GMP19} that the interacting system is gapped ({\it i.e.}, the Euclidean correlation functions decay exponentially) for {\it all} the values of $(\phi,M)$ outside
a pair of {\it renormalized critical lines}, provided $\mu$ is chosen
appropriately. The shape of the renormalized critical lines is
qualitatively the same as that in {\it Fig. \ref{fig.3}} and the values of $\sigma_{12}$ inside
or outside the curves are the same as in the non-interacting setting. The proof in \cite{GJMP,GMP19} is valid arbitrarily close to the renormalized critical lines, that is, uniformly in the gap (which vanishes as the critical lines are approached). The generalization of Theorem 1 to this setting requires an infrared multiscale analysis, reminiscent of the one developed in \cite{GM} for a model of graphene with short range interactions. 

\medskip

\noindent4) The methods reviewed here can be used to show that, under the gap condition, for $U$ sufficiently small, the interacting Hamiltonian has a spectral gap above the unique ground state, uniformly in the 
volume $L$. For a recent independent proof of this fact, see \cite{Has}. Knowing this fact, one can show that the interacting Hall conductivity $\sigma_{12}$ has an interpretation in terms of a geometrical 
index  \cite{AS,HM}, analogous to the representation in terms of the first Chern number of the Bloch bundle that is valid for the non-interacting one. Note, however, that such a 
representation involves a many-body (rather than a one-body) projector, that is the projector $\Pi$ on the interacting, many-body, ground state. The control parameter one needs to average over
is not the quasi-momentum $k$ (which is not a priori well-defined in the many-body case), as in \eqref{eq.2}, but, rather, an angle $\theta$ controlling the (twisted) boundary conditions. The geometrical 
representation underlying the results in \cite{AS,HM} is not suitable for direct analytical computations: therefore, while the results in \cite{AS,HM}
show that the Hall conductivity is quantized, they do not provide one with a good representation formula for computing the explicit value of the Hall conductivity in given models. On the contrary, the approach reviewed in this notes
provides a very simple explicit formula for such a value.

\section{Overview of the proof} Let me now start to discuss the main strategy to be followed for proving the main theorem. In order to compute the Kubo conductivity \eqref{eq.0}, the key point is to 
control the current-current correlations, which enter the first term in the right side of \eqref{eq.0}. Actually, the direct control of the real-time interacting correlations $\langle\!\!\langle [J_i,J_j(t)]\rangle\!\!\rangle_{\infty}$ is very difficult. In order to go around the obstacle, one can take advantage of the powerful methods (applicable both in the massive and in the massless cases), based on ideas and techniques of 
{\it constructive field theory}, which can be used to control and compute the {\it Euclidean, imaginary-time}, current-current correlations. 
These methods will be briefly reviewed in the following. Before discussing them, let me explain how to use informations on the Euclidean correlations in order to infer informations on the Kubo conductivity. 

\subsection{Euclidean correlations}\label{sec.5.1} Let $O^{(1)},
\ldots, O^{(n)}$ be observables, each of which is assumed to be a
linear combination of normal-ordered, even, monomials in the fermionic
creation/ annihilation operators. I let 
$$O^{(i)}_t\equiv O^{(i)}(-it)=e^{tH}O^{(i)}e^{-tH}$$ 
be the imaginary-time evolution of the $i$-th observable, with $0\le t<\beta$, and 
$$\langle T\, O^{(1)}_{t_1}\cdots O^{(n)}_{t_n}\rangle_{\beta,L}=\frac{{\rm Tr}\, e^{-\beta H} O^{(1)}_{t_1}\cdots O^{(n)}_{t_n}}{{\rm Tr}\, e^{-\beta H}}$$
where $T$ is the (multilinear) time-ordering operator, acting on monomials in the fields $\psi^\pm_{(t,x)\sigma}=e^{tH}\psi^\pm_{x,\sigma}e^{-tH}$ as: 
$$T\, \psi^{\epsilon_1}_{(t_1,x_1)\sigma_1}\cdots \psi^{\epsilon_n}_{(t_n,x_n)\sigma_n}=(-1)^{\pi}\psi^{\epsilon_{\pi_1}}_{(t_{\pi_1},x_{\pi_1})\sigma_{\pi_1}}\cdots \psi^{\epsilon_{\pi_n}}_{(t_{\pi_n},x_{\pi_n})\sigma_{\pi_n}}$$
where $\pi$ is the permutation of $(1,\ldots,n)$ such that $t_{\pi_1}>\cdots>t_{\pi_n}$ (in case of coinciding times, the variables at equal time will be assumed to be normal ordered, after the action of  the ordering operator $T$). I also indicate by $\langle T\, O^{(1)}_{t_1};\cdots ; O^{(n)}_{t_n}\rangle_{\beta,L}$ the corresponding cumulants, or truncated correlations, {\it i.e.}, 
$\langle T\, O^{(1)}_{t_1};O^{(2)}_{t_2}\rangle_{\beta,L}=\langle T\, O^{(1)}_{t_1}O^{(2)}_{t_2}\rangle_{\beta,L}-\langle O^{(1)}\rangle_{\beta,L}\langle O^{(2)}\rangle_{\beta,L}$, etc. 
Moreover, at finite $\beta$, I introduce the notion of Fourier transform with respect to the Euclidean time: 
$$\hat O_\omega=\int_0^\beta dt\, e^{-i\omega t}O_t, $$
with $\omega\in\frac{2\pi}{\beta}{\mathbb Z}$ the Matsubara frequency. Note that the following momentum conservation rule holds: 
\begin{eqnarray} &&\int_0^{\beta}dt_1\cdots\int_0^{\beta}dt_n\langle T\, O^{(1)}_{t_1};\cdots ; O^{(n)}_{t_n}\rangle_{\beta,L}\,e^{-i\omega_1t_1\cdots-i\omega_nt_n}=\nonumber\\
&&\hskip5.truecm =\delta_{\omega_1+\cdots+\omega_n,0}
\langle T\, \hat O^{(1)}_{\omega_1};\cdots ; \hat O^{(n)}_{\omega_n}\rangle_{\beta,L}.\nonumber\end{eqnarray}

\bigskip

A natural question arises: why are these definitions interesting and useful? The main reason is that the Euclidean correlation functions appear naturally in the perturbation theory 
for the equilibrium correlations, via the use of Trotter's product formula and the Duhamel expansion. In many cases, there are powerful methods (fermionic Renormalization Group) for controlling 
the interacting equilibrium correlation functions at weak enough
coupling, and to derive detailed informations about their analyticity properties. 

\medskip

\begin{figure}[ht]
\centering
\includegraphics[width=0.55\textwidth]{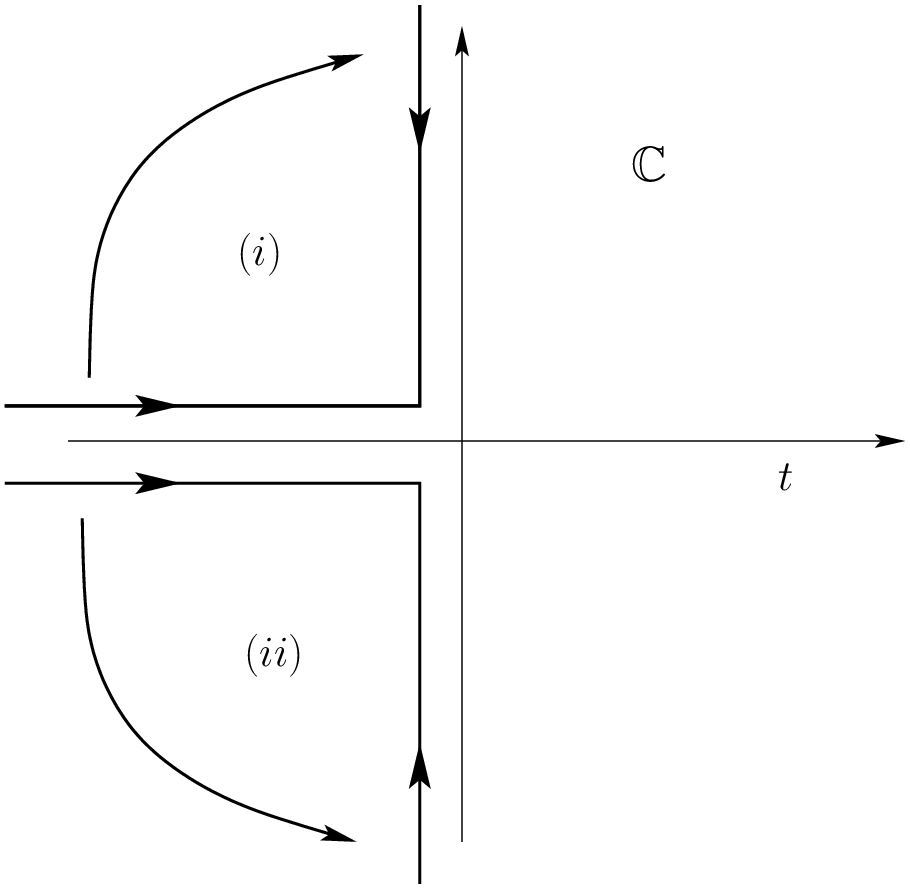}
\caption{\it The `complex rotations' of the integration paths that we perform on the two contributions (i) and (ii) to the Kubo conductivity. }
\label{fig.4}
\end{figure}

Moreover, typically, once one knows how to control the equilibrium correlations, one also knows how to control the interacting, Euclidean-time, correlations, including control of their analyticity properties in the 
complex time plane. Finally, if one can prove that the complex time plane is free of singularities of the Euclidean correlation, one can a posteriori reconstruct the (integral of the) real-time ones via a `Wick rotation' in the complex time plane. In order to give an idea of what I am referring to, let me consider our main object of interest, {\it i.e.}, the Kubo conductivity eq.\eqref{eq.0}, and let me focus on the first term in the right side,
$\frac{-i}{\eta}\int_{-\infty}^0dt\, e^{\eta t}\pmb{\langle}
[J_i,J_j(t)]\pmb{\rangle}_\infty$: this can be thought of as the sum
of two terms: (i) $\frac{-i}{\eta}\int_{-\infty}^0dt\, e^{\eta
  t}\pmb{\langle} J_iJ_j(t)\pmb{\rangle}_\infty$, and (ii)
$\frac{i}{\eta}\int_{-\infty}^0dt\, e^{\eta t}\pmb{\langle} J_j(t)
J_i\pmb{\rangle}_\infty$. Now, if the two terms are free of
singularities in the second and third quadrants of the complex plane,
respectively, then the two integrals can be rotated as shown in {\it Fig. \ref{fig.4}}, provided the contributions from the integrals over the two quarter circles at infinity give vanishing contribution. 

After having performed the rotation, the sum of the two terms can be rewritten as $-\frac1{\eta}\hat K_{ij}(-\eta)$, where
\begin{equation}\hat K_{ij}(\omega)=\lim_{\beta\to\infty}\lim_{L\to\infty}\frac1{\beta L^2}\langle T\, \hat J_{i,\omega}; \hat J_{j,-\omega}\rangle_{\beta,L}.\label{Komega}\end{equation}
Moreover, one can check that the diamagnetic contribution to the Kubo conductivity can be rewritten as $\hat K_{ij}(0)/\eta$. 
As a result, the Kubo conductivity is equal to the Euclidean one: $\sigma_{ij}=\bar\sigma_{ij}$, with 
\begin{equation}\bar\sigma_{ij}=\frac{1}{|\ell_1\wedge\ell_2|}\lim_{\omega\to 0^-}\frac{\hat K_{ij}(\omega)-\hat K_{ij}(0)}{\omega}.\label{barsigma}\end{equation}
At this point, I am finally in the position of describing the general strategy of the proof of the main theorem: 

\medskip

\noindent1) One first discusses the existence of the thermodynamic and zero temperature limit of the Euclidean correlations, and proves their analyticity in $U$ and in $\omega$ in suitable regions of the complex plane.
Using the analyticity of the Euclidean correlations in the complex $t$-plane, as well as the existence of the real-time correlations in the 
$\beta,L\to\infty$ limit (which follows from the Lieb-Robinson bounds \cite{LR}), one proves that the `Wick rotation strategy' 
sketched above is rigorously justified, thus showing that the Kubo conductivity is equal to its Euclidean counterpart. 
%Note that, by the analyticity of the Euclidean correlations, the Euclidean Kubo conductivity is equal to the sum of its Taylor series.

\medskip

\noindent2) 
%In light of the previous item, the claim of the main theorem 
%is equivalent to the claim that all the Taylor coefficients of the Euclidean Kubo conductivity beyond the $0$-th order one vanish. 
In order to show that the interacting Euclidean Kubo conductivity equals the non-interacting one, 
one combines two classes of remarkable cancellations/identities satisfied by the Euclidean correlation functions, known as Ward Identities (more precisely, the Ward Identities associated with the 
continuity equation for the current) and Schwinger-Dyson equations. {\it Note}: the idea of using these two classes of identities to 
prove the vanishing of the interaction corrections to the Euclidean conductivity is due to \cite{CH}, whose strategy of proof is adapted here to the case of the Hall conductivity of interacting lattice fermions. 

\medskip

In the following sections, I discuss these two steps in some detail. 

\section{Equilibrium perturbation theory}

Let me start, for simplicity, by discussing the perturbation theory for the partition function of the system at finite $\beta,L$: 
$${\rm Tr}\, e^{-\beta H}={\rm Tr}\, e^{-\beta (H_0-\mu N+UV)}.$$
Since the fermionic 
operators are all bounded at finite $L$, one can rewrite this trace by using the Duhamel's expansion (see, {\it e.g.}, \cite[Section 3.2]{Ue}), thus getting: 
%\begin{eqnarray} && {\rm Tr}e^{-\beta H}={\rm Tr}e^{-\beta (H_0-\mu N)}\, \Big[ 1+\sum_{n\ge 1}(-U)^n\int_{\beta>t_%1>\cdots>t_n>0} \hskip-1.3truecm dt_1\cdots dt_n\, \langle V_{t_1}\cdots V_{t_n}\rangle^0_{\beta,L}\Big]\nonumber\\
%&&={\rm Tr}e^{-\beta (H_0-\mu N)}\, \Big[ 1+\sum_{n\ge 1}\frac{(-U)^n}{n!}\int_0^\beta dt_1\cdots\int_0^\beta dt_n %\langle T\,  V_{t_1}\cdots V_{t_n}\rangle^0_{\beta,L}\Big],\label{eq.6.1}\end{eqnarray}
\begin{equation}
\vcenter{\hsize=\hsize\openup1\jot\halign{
\hfil$\displaystyle{#}$
&$\displaystyle{#}$\hfil
\cr
{\rm Tr}e^{-\beta H}
&={\rm Tr}e^{-\beta (H_0-\mu N)}\,\times\cr
&\quad \Big[ 1+\sum_{n\ge 1}(-U)^n\int_{\beta>t_1>\cdots>t_n>0} \hskip-1.3truecm dt_1\cdots dt_n\, \langle V_{t_1}\cdots V_{t_n}\rangle^0_{\beta,L}\Big]\cr
& ={\rm Tr}e^{-\beta (H_0-\mu N)}\,\times\cr
&\quad \Big[ 1+\sum_{n\ge 1}\frac{(-U)^n}{n!}\int_0^\beta dt_1\cdots\int_0^\beta dt_n \langle T\,  V_{t_1}\cdots V_{t_n}\rangle^0_{\beta,L}\Big],
\cr
}}\label{eq.6.1}
\end{equation}
where $V_t=e^{t(H_0-\mu N)}Ve^{-t(H_0-\mu N)}$ and $\langle\cdot\rangle^0_{\beta,L}$ is the finite volume, finite temperature, Gibbs average with respect to the Hamiltonian $H_0-\mu N$.

At finite volume, the series in $U$ for $Z(U): ={\rm Tr}e^{-\beta H}$ is convergent (even more: the function $Z(U)$ is {\it entire} in $U$, for any finite $L$). The series for the logarithm of $Z(U)$ is a priori 
convergent in a very small neighborhood of the origin. In fact,
$$Z(U)\ge 1-\sum_{n\ge 1}\frac{1}{n!}(\beta|U|)^n\|V\|_\infty^n=2-e^{\beta|U|\,\|V\|_\infty},$$
which does not vanish if $U\in B:=\{U\in \mathbb C: \ |U|< \frac{\log 2}{\beta\|V\|_\infty}\}$. Therefore, the series of $\log Z(U)$ is convergent in any compact set $K$ belonging to the open ball $B$.
Note that typically $\|V\|_\infty$ is 
proportional to $L^2$, so that the radius of $B$ goes to zero as $L\to\infty$. The goal of the following discussion is to show that the analyticity region of
$\log Z(U)$ can be extended to a ball centered at the origin, with radius {\it independent} of $L$. 

Before I proceed with the discussion, let me note that similar considerations are valid for the series in $U$ for the Euclidean correlations: if $\beta>t_1>t_2>0$, the Euclidean correlation $\langle O^{(1)}_{t_1}
O^{(2)}_{t_2}\rangle_{\beta,L}$ is equal to the ratio between Tr$e^{-\beta H}e^{t_1H}O^{(1)}e^{-(t_1-t_2)H}O^{(2)}e^{-t_2H}$ and $Z(U)$. The denominator is expanded as described above, and the numerator 
admits a similar expansion: 
$$
\displaylines{
\langle O^{(1)}_{t_1}O^{(2)}_{t_2}\rangle_{\beta,L}=\langle
O^{(1)}_{t_1}O^{(2)}_{t_2}\rangle_{\beta,L}^0
\hfill\cr\hfill
+\sum_{n\ge 1}\frac{(-U)^n}{n!}\int_0^\beta ds_1\cdots\int_0^\beta ds_n
\langle T\,  O^{(1)}_{t_1}O^{(2)}_{t_2}V_{s_1}\cdots V_{s_n}\rangle_{\beta,L}.
}
$$
Therefore, $\langle T\, O^{(1)}_{t_1}
O^{(2)}_{t_2}\rangle_{\beta,L}$ is the ratio of two entire functions, the denominator being different from zero in the analyticity domain of $\log Z(U)$.

\begin{figure}[ht]
\centering
\includegraphics[width=0.3\textwidth]{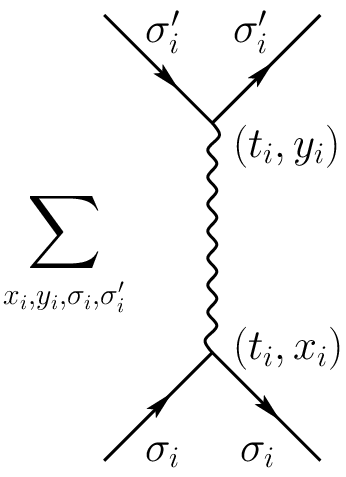}
\caption{\it The graphical representation of $V_{t_i}$ as a `four-legged' vertex. The wiggly line represents the interaction $v_{\sigma_i\sigma'_i}(x_i-y_i)$, 
while the oriented solid lines (the `external legs') represent the operators $\psi^\pm_{(t_i,x_i)\sigma_i}$ and $\psi^\pm_{(t_i,y_i)\sigma_i'}$, with the convention that the entering (resp. exiting) lines represent the annihilation (resp. creation) operators $\psi^-$ (resp. $\psi^+$).}
\label{fig.5}
\end{figure}

Now, how do we evaluate $\log Z(U)$, and how do we prove bounds on its analyticity domain? The starting point is an explicit representation of the $n$-th Taylor coefficient of $Z(U)$: from \eqref{eq.6.1}
this coefficient is $(n!)^{-1}$ times the integral over $t_1\cdots t_n$ of $\langle T\,  V_{t_1}\cdots V_{t_n}\rangle^0_{\beta,L}$. Such an average can be computed via the fermionic Wick's rule, which is readily 
explained in terms of the following graphical
representation. Associate with every
$$
V_{t_i}=\sum_{x_i,y_i,\sigma_i,\sigma_i'}
\psi^+_{(t_i,x_i)\sigma_i}\psi^-_{(t_i,x_i)\sigma_i}v_{\sigma_i\sigma'_i}(x_i-y_i)
\psi^+_{(t_i,y_i)\sigma'_i}\psi^-_{(t_i,y_i)\sigma'_i}
$$
the `four-legged' vertex in {\it Fig. \ref{fig.5}}.

\begin{figure}[ht]
\centering
\includegraphics[width=0.95\textwidth]{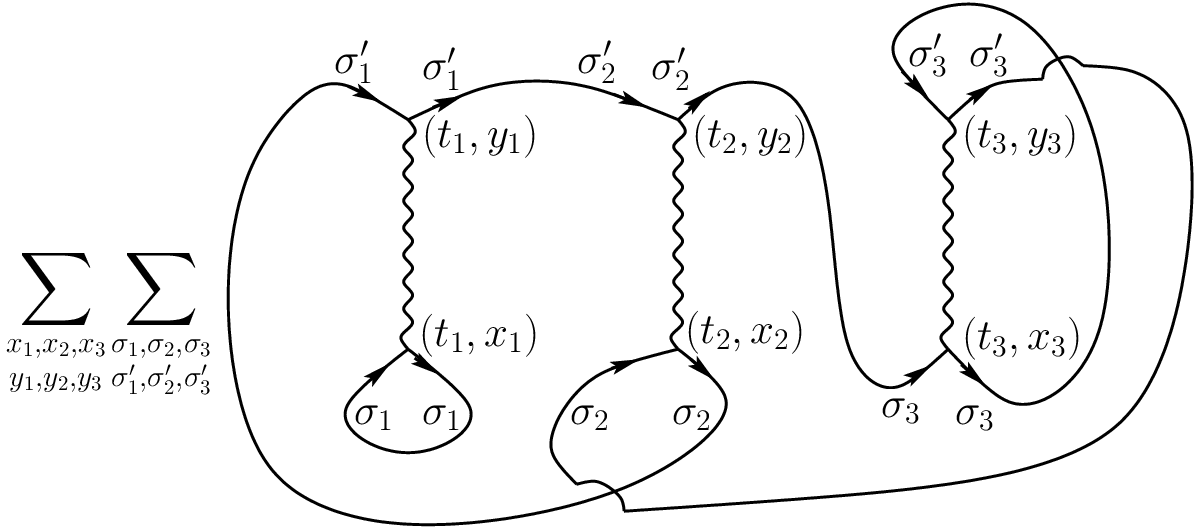}
\caption{\it An example of a Feynman diagram of order $n=3$. The solid lines are obtained by the pairing, or contraction, of two of the external legs 
of the vertices under consideration. Any pair of contracted legs must have compatible orientation, {\it i.e.}, an exiting external leg can only be contracted with an entering one. Each solid line obtained by the contraction of two legs corresponds to a propagator, with indices readable from the labels of the two contracted legs.}
\label{fig.6}
\end{figure}

Draw $n$ such vertices and pair their `external legs' in all possible ways , in such a way that 
the orientations of the paired lines are compatible two by two. The graph associated with any such pairing is called a `Feynman diagram', and 
$\langle T\,  V_{t_1}\cdots V_{t_n}\rangle^0_{\beta,L}$ can be
expressed as the sum over all the possible Feynman diagrams of their
values, which are defined as follows. For an example of a Feynman
diagram of order $n=3$, see {\it Fig. \ref{fig.6}}.

First of all, every Feynman diagram $\mathcal G$ comes with a permutation $\pi=\pi(\mathcal G)$, associated with the corresponding pairing of fermionic fields, which is the one 
needed to move every creation operator $\psi^+_{(t',x')\sigma'}$ immediately to the right of the annihilation operator $\psi^-_{(t,x)\sigma}$, which it is connected with. Moreover, 
every contracted pair $(\psi^-_{(t,x)\sigma},\psi^+_{(t',x')\sigma'})$ comes with the value 
\begin{eqnarray} &&g_{\sigma\sigma'}^{\beta,L}(x-x',t-t')=\langle T\, \psi^-_{(t,x)\sigma}\psi^+_{(t',x')\sigma'}\rangle_{\beta,L}^0=
\frac1{L^2}\sum_{k\in\mathcal B_L}e^{-ik(x-x')}\times\nonumber\\
&&\qquad \times \Big\{e^{-(\hat H^0(k)-\mu)(t-t')}\Big[\frac{\mathds 1_{t>t'}}{1+e^{-\beta(\hat H^0-\mu)}}-
\frac{\mathds 1_{t\le t'}}{1+e^{\beta(\hat H^0-\mu)}}\Big]\Big\}_{\sigma\sigma'},\nonumber\end{eqnarray}
which is called the `propagator'. Note that its $\beta,L\to\infty$ limit is 
$$g_{\sigma\sigma'}(x,t)=\int_{\mathcal B}\frac{dk}{|\mathcal B|}e^{-ikx}\{e^{-(\hat H^0(k)-\mu)t}[\mathds 1_{t>t'}P_+(k)-\mathds 1_{t\le t'}P_-(k)]\}_{\sigma\sigma'}.$$
In the following, I will use the shorthand notation $g_\ell$ for the propagator associated with the pair $\ell=\big(((t,x),\sigma), ((t',x'),\sigma') \big)$. I also denote by $G(\underline x, \underline y,$ $\underline t,$  
$\underline\sigma, \underline\sigma')$
the set of Feynman diagrams obtained from the contraction of the quartic monomials labeled by the time/space/color variables\break $\{\big(t_i,(x_i,y_i),(\sigma_i,\sigma_i')\big)\}_{i=1,\ldots,n}$ (here 
I used the symbol $\underline x$ to denote the $n$-ple $(x_1,\ldots,x_n)$, etc.)

Given these definitions, the value Val$(\mathcal G)$ of the Feynman diagram $\mathcal G\in G(\underline x, \underline y, \underline t, 
\underline\sigma, \underline\sigma')$ is 
\begin{equation}{\rm Val}(\mathcal G)=(-1)^\pi\prod_{i=1}^n v_{\sigma_i\sigma_i'}(x_i-y_i)\prod_{\ell\in\mathcal G}g_\ell\label{-1pi}\end{equation}
and
$$\langle T\,  V_{t_1}\cdots V_{t_n}\rangle^0_{\beta,L}=\sum_{\substack{\underline x,\underline y\\ \underline\sigma,\underline\sigma'}}
\hskip.2truecm
\sum_{\mathcal G\in G(\underline x, \underline y, \underline t, 
\underline\sigma, \underline\sigma')} {\rm Val}(\mathcal G),$$
so that 
$$Z(U)=1+\sum_{n\ge 1}\frac{(-U)^n}{n!}\int_0^\beta dt_1\cdots\int_0^\beta dt_n 
\, \sum_{\substack{\underline x,\underline y\\ \underline\sigma,\underline\sigma'}}
\hskip.2truecm
\sum_{\mathcal G\in G(\underline x, \underline y, \underline t, 
\underline\sigma, \underline\sigma')} {\rm Val}(\mathcal G).$$
A remarkable combinatorial theorem, known as the linked cluster theorem \cite{AGD,NO}, shows that the logarithm of the partition function 
admits a very similar expansion, with the important difference that the Feynman diagrams contributing to $\log Z(U)$ are all and only the {\it connected} diagrams in\break 
$ G(\underline x, \underline y, \underline t, 
\underline\sigma, \underline\sigma')$. A similar discussion is valid for the correlation functions: in particular, truncated correlations are expressed as sums
of values of connected Feynman diagrams, with vertices corresponding both to the quartic interactions introduced above and to extra vertices corresponding to the insertion 
of the operators one is computing the correlations of. 
%In order not to overwhelm the discussion, we will not belabor this aspect further in these notes.
In conclusion,
\medium
$$\log Z(U)=\sum_{n\ge 1}\frac{(-U)^n}{n!}\int_0^\beta dt_1\cdots\int_0^\beta dt_n 
\, \sum_{\substack{\underline x,\underline y\\ \underline\sigma,\underline\sigma'}}
\hskip.2truecm
\sum_{\mathcal G\in G_{conn}(\underline x, \underline y, \underline t, 
\underline\sigma, \underline\sigma')} {\rm Val}(\mathcal G),$$
\normalsize
where
\medium $G_{conn}(\underline x, \underline y, \underline t, 
\underline\sigma, \underline\sigma')$\normalsize
is the set of
connected diagrams in
\medium$ G(\underline x, \underline y, \underline t, 
\underline\sigma, \underline\sigma')$.\normalsize

Now, how do we estimate the $n$-th order contribution to this series? The basic object one needs to evaluate is the integral over the time, space and color 
variables of a generic connected Feynman diagram with $n$ quartic vertices, {\it i.e.}, $\int \sum \big|{\rm Val}(\mathcal G)\big|$, where the integral is over the time variables, and the sum over the space and color 
indices. Recalling that $\big|{\rm Val}(\mathcal G)\big|=\prod_{i}|v_{\sigma_i\sigma_i'}(x_i-y_i)|\prod_{\ell\in\mathcal G}|g_\ell|$, one can evaluate the desired integral by selecting a minimal connecting 
subset of the set of wiggly and solid lines, which the Feynman consists of (this minimal connecting subset is by construction a tree, to be called a `spanning tree'). I choose the spanning tree in such a way that 
it contains all the wiggly lines of the diagram, and a suitably chosen set of $n-1$ solid lines. I bound the propagators on the solid lines outside the spanning tree by their $L_\infty$ norm; once this is done, 
the integration over the space and time variables of the product of the interactions and propagators on the spanning tree produces the product of their $L_1$ norms, so that, recalling that $|I|$ is the cardinality of the set of allowed colors, 
$$\int \sum \big|{\rm Val}(\mathcal G)\big|\le \beta L^2\|v\|_1^n |I|^{2n} \|g\|_1^{n-1}\|g\|_\infty^{n+1},$$
where
$$
\|g\|_\infty=\sup_{\s,\s'}\sup_{x,t}|g_{\sigma\sigma'}^{\beta,L}(x,t)|\ ,\quad 
\|g\|_1=\sup_{\s,\s'}\int dt \sum_x|g_{\sigma\sigma'}^{\beta,L}(x,t)|,
$$
and similarly for $\|v\|_1$. 
Recall that $v$ is of finite range, so its $L_1$ norm is finite. Moreover, under the gap condition, the propagator decays exponentially in space-time, uniformly in $\beta,L$: therefore, both 
its $L_1$ and $L_\infty$ norms are finite, uniformly in the temperature and volume. In conclusion, the $n$-th order term in the series of $\frac1{\beta L^2}\log Z(U)$ is bounded by 
$$\frac{C^n}{n!}|U|^n\{\#\ {\rm Feynman}\ {\rm diagrams}\ {\rm of}\ {\rm order}\ n\},$$
for a suitable constant $C$. Now, the number of Feynman diagrams of order $n$ equals $(4n-1)!!$, which grows like $({\rm const.})^n(n!)^2$ as\footnote{By taking into account the connectedness
condition, one could improve the bound above, by replacing  $\{\#\ {\rm Feynman}\ {\rm diagrams}\ {\rm of}\ {\rm order}\ n\}$ by $\{\#\ {\rm connected}\ {\rm Feynman}\ {\rm diagrams}\ {\rm of}\ {\rm order}\ n\}$.
However, one can easily convince oneself that both quantities grow like $({\rm const.})^n(n!)^2$ for large $n$, so the connectedness condition does not qualitatively change things.}
$n\to\infty$ and, therefore, the bound that we just derived is not summable over $n$. In other words, {\it the series of Feynman diagrams is not absolutely convergent}. 

Remarkably, it is possible to properly resum the series of connected Feynman diagrams, by carefully taking into account the $(-1)^\pi$ signs in \eqref{-1pi}, in such a way that the resummed series is convergent. 
The basic observation is that 
$$
\displaylines{\qquad
\langle T\,
\psi^-_{(t_1,x_1)\sigma_1}\psi^+_{(t'_1,y_1)\sigma_1'}\cdots
\psi^-_{(t_n,x_n)\sigma_n}\psi^+_{(t'_n,y_n)\sigma_n'}\rangle^0_{\beta,L}=
\hfill\cr\hfill
\det\big[g_{\sigma_i,\sigma_j'}(x_i-y_j,t_i-t_j')\big]_{i,j=1,\ldots,n}.
\qquad}
$$
By expanding the determinant, one obtains $n!$ terms, each of which corresponds to one specific pairing of the fermionic fields (i.e, to one `Feynman diagram'); each of these terms is bounded by 
$\|g\|_\infty^n$, so this expression can be bounded by $\|g\|_\infty^n n!$. However, one can do much better than this: if instead of expanding the determinant, one recalls that it is equal to the product of its 
eigenvalues, one sees that its absolute value typically scales like $({\rm const.})^n$, which is combinatorially better by a factor $1/n!$. This observation suggests that by grouping the 
sum over connected Feynman diagrams in determinant form (which should be possible, thanks to the $(-1)^\pi$ signs coming from the fermionic statistics), then one should be able to improve the 
bound derived above by a factor $({\rm const.})^n/n!$. The rough idea is to group together the connected Feynman diagrams sharing a common spanning tree, and then to resum the result of the pairings outside 
the spanning tree in the form of a determinant, to be bounded by $({\rm const.})^n$. Since the number of spanning trees over $n$ vertices is of the order $({\rm const.})^n n!$ for large $n$, this procedure
seems in fact to improve the previous bound by the desired combinatorial factor. 

Of course, one needs to proceed with care, in order to avoid overcounting of the diagrams, when summing over the spanning trees. The key 
combinatorial formula was proposed long ago by Brydges, Battle and Federbush \cite{BF, Br, BrF}, later improved in collaboration with Kennedy \cite{BK}, and reads as follows:
let $\Psi(P_i)$ be a shorthand for the quartic monomial associated
with the $i$-th quartic vertex, {\it i.e.},
$$
\Psi(P_i)=
\psi^+_{(t_i,x_i)\sigma_i}\psi^-_{(t_i,x_i)\sigma_i}v_{\sigma_i\sigma'_i}(x_i-y_i)
\psi^+_{(t_i,y_i)\sigma'_i}\psi^-_{(t_i,y_i)\sigma'_i}\>,
$$
with $P_i$ the set of four labels (to be called `field labels') associated with the four 
creation/annihilation operators, then 
\begin{equation}
  \vcenter{\openup1\jot\halign{
      \hfil$\displaystyle{#}$
      &$\displaystyle{#}$\hfil
      \cr
      \langle \Psi(P_1);\cdots;&\Psi(P_n)\rangle_{\beta,L}^0
      \cr
 &     =\sum_{T}\alpha_T\prod_{\ell\in T}g_\ell
\int dP_T(\underline s)\det[s_{i(f),i(f')}g_{(f,f')}],
      \cr
      }}
\label{eq.6.2}\end{equation}
where the sum is over the collections $T$ of $n-1$ propagators (graphically, of solid lines, each corresponding to the contraction of a pair of field labels $f_1,f_2\in\cup_iP_i$) guaranteeing minimal connection among the $n$ vertices
({\it i.e.}, the union of $\,T$ and of the wiggly lines of the $n$ vertices forms a spanning tree), 
$\alpha_T$ is a sign (irrelevant for the purpose of proving convergence of the series), and $dP_T(\underline s)$ is a probability measure, with the variables $\underline s=(s_{ij})_{i,j=1,\ldots n}$ playing the role of interpolation parameters, supported on a set of $s_{ij}$'s such that 
$s_{ij}=(u_i,u_j)$, for a family of vectors $u_i\in \mathbb C^n$ of unit norm. Finally, the matrix $[s_{i(f),i(f')}g_{(f,f')}]$ has elements labelled by field labels
$f,f'\in\{\cup_iP_i\}\setminus P_T$, where $P_T$ is the set of field labels corresponding to the propagators in $T$ ({\it i.e.}, $[s_{i(f),i(f')}g_{(f,f')}]$ is labelled by fields 
outside the spanning tree); in writing $s_{i(f),i(f')}$, $i(f)$ indicates the vertex which the field labelled by $f$ belongs to; moreover, 
$g_{(f,f')}$ is a shorthand for $g_{\sigma(f),\sigma(f')}(x(f)-x(f'),t(f)-t(f'))$.

In order to use effectively the `BBFK interpolation formula', I must discuss how to bound the determinant. Life is easy if $g_{(f,f')}$ is in Gram form, {\it i.e.}, if there exist vectors of finite norm in a separable Hilbert space, such that 
$g_{(f,f')}=(a_f,b_{f'})$, in which case (Gram-Hadamard inequality, see, {\it e.g.}, \cite[Appendix A.3]{GeM})
$$\det [s_{i(f),i(f')}g_{(f,f')}]=\det(u_{i(f)}\otimes a_f,u_{i(f')}\otimes b_{f'})\le \prod_f\|a_f\|\, \|b_f\|.$$
In our case the propagator is not in Gram form, but almost so (we give here the formulas in the $\beta,L\to\infty$ limit; similar, but slightly more cumbersome, expressions are valid in the general case): 
$$g_{(f,f')}
=\mathds 1_{t(f)>t(f')} A^+_{f,f'}-\mathds 1_{t(f)\le t(f')}A^-_{f,f'},$$
where 
$$A^\pm_{f,f'}=\int_{\mathcal B}\frac{dk}{|\mathcal B|}e^{-ik(x(f)-x(f'))}\big[e^{-(\hat H^0(k)-\mu)(t(f)-t(f'))}P_\pm(k)\big]_{\sigma(f),\sigma(f')}.$$
Note that both $A^+_{f,f'}$ and $A^-_{f,f'}$ are in Gram form: in fact, 
\begin{eqnarray}A^\pm_{f,f'}\!\!\!&=&\!\!\!\iint\limits_{\mathcal B\times \mathbb R}\frac{dk d\omega}{|\mathcal B|\pi} \sum_{s\in I}\big[a^\pm_{x(f),t(f)\sigma(f)}(k,\omega,s)\big]^*a^\pm_{x(f'),t(f'),\sigma(f')}(k,\omega,s)\nonumber\\
&\equiv&\Big( a^\pm_{x(f),t(f),\sigma(f)}, a^\pm_{x(f'),t(f'),\sigma(f')}\Big),\label{inner}\end{eqnarray}
where 
$$a^\pm_{x,t,\sigma}(k,\omega,s)=e^{ik x+i\omega t}\Biggl[\sqrt{\pm(\hat H^0(k)-\mu)}\frac{P_\pm(k)}{-i\omega+\hat H^0(k)-\mu}\Biggr]_{s,\sigma}.$$
From the definitions, letting $\|\cdot\|$ be the norm induced by the inner product in \eqref{inner}, one finds
$$\|a_{x,t,\sigma}^\pm\|^2=\int\frac{dk}{|\mathcal B|}\Big[P_\pm(k)\Big]_{\sigma,\sigma},$$
which is $\le 1$, for both signs $\pm$. Summarizing, the determinant in the right side of \eqref{eq.6.2} admits the following `time-ordered' Gram 
representation:
%\begin{eqnarray}&&\det[s_{i(f),i(f')}g_{(f,f')}]=\label{eq.6.3}\\
%&&\hskip-.7truecm= 
%\det\big[\mathds 1_{t(f)>t(f')} \big(u_{i(f)}\otimes a^+_{f},u_{i(f')}\otimes a^+_{f'}\big)
%-\mathds 1_{t(f)\le t(f')}\big(u_{i(f)}\otimes a^-_{f},u_{i(f')}\otimes a^-_{f'}\big)\big],\nonumber\end{eqnarray}
%%
\begin{eqnarray}
  \vcenter{\openup1\jot\halign{
      \hfil$\displaystyle{#}$
      &$\displaystyle{#}$\hfil
      \cr
\det[s_{i(f),i(f')}g_{(f,f')}]&=\label{eq.6.3}
\det\big[\mathds 1_{t(f)>t(f')} \big(u_{i(f)}\otimes a^+_{f},u_{i(f')}\otimes a^+_{f'}\big)
\cr
&\qquad -\mathds 1_{t(f)\le t(f')}\big(u_{i(f)}\otimes a^-_{f},u_{i(f')}\otimes a^-_{f'}\big)\big],
      \cr
      }}
\end{eqnarray}
which is in fact enough for our purposes: as shown in \cite{BrP2,PS08}, the right side of \eqref{eq.6.3} is bounded above by 
$(\|a^+\|+\|a^-\|)^{2d}$, where $d$ is the dimension of the matrix (in our case, $d=n+1$). In conclusion, recalling that $\|a_{x,t}^\pm\|\le 1$, 
one finds that the integral of the determinant in the right side of \eqref{eq.6.2} is bounded from above by $2^{2(n+1)}$. Recalling also that the number of 
collections $T$ (`number of spanning trees') is bounded above by $({\rm const.})^nn!$, one sees that this estimate is enough for our purposes, {\it i.e.}, 
to obtain uniform convergence of the specific free energy, for $U$ small enough. In fact, in view of the combinatorial gain obtained above, the 
$n$-th order term in the series for $\log Z(U)$ is bounded above by $C^n|U|^n$, for a suitable constant $C>0$, which is summable over $n$ for $|U|<C^{-1}$.

A similar argument can be repeated to prove the convergence of the series in $U$ for the Euclidean truncated correlation functions of local operators in the same 
domain of analyticity, as well as their exponential decay at large space-time distances, uniformly in $\beta$ and $L$. 
In particular, there exists $C,c>0$ such that  
\begin{equation}\pmb{\langle} J_{i,t} ;J_j\pmb{\rangle}_\infty\le C e^{-c\delta_\mu t}, \qquad \forall t\ge 0. \label{eq.td}\end{equation}
The details of the proof will not be belabored here and are left as an instructive 
exercise to the reader.

\subsection{Wick rotation for the Kubo conductivity}

Using the analyticity and exponential decay in space-time of the Euclidean correlations discussed above, one can prove that the 
Wick rotation for the Kubo conductivity, sketched in Section \ref{sec.5.1}, can be made rigorous. The key steps in the proof 
are the following (for details, consult \cite[Section 6]{GMP17}):

\medskip

\noindent1) Let $J_i(z)=e^{izH}J_ie^{-izH}$, $z\in\mathbb C$, be the complex time evolution of the current. Note that $J_{i,t}=J_{i}(-it)$, $t\in\mathbb R$. 
At finite $\beta,L$, the correlation 
$L^{-2}\langle J_i(z);J_j\rangle_{\beta,L}$ is entire in $z$. Moreover, thanks to the results on the Euclidean correlations sketched above, 
these finite volume correlations converge as $\beta,L\to\infty$ along the negative imaginary axis, and they are uniformly bounded in the whole complex 
lower half-plane. Therefore, by Vitali's theorem, the limit as $\beta,L\to\infty$ of the complex time correlations,
$\pmb{\langle}  J_i(z);J_j\pmb{\rangle}_\infty$ exists and is analytic in the 
whole open lower half-plane. Moreover, by Cauchy-Schwarz inequality and \eqref{eq.td}, 
$$|\pmb{\langle} J_i(z);J_j\pmb{\rangle}_\infty|\le \pmb{\langle}  J_i({\rm Re}\,z);J_i \pmb{\rangle}_\infty^{1/2}
 \pmb{\langle}  J_j({\rm Re}\,z);J_j \pmb{\rangle}_\infty^{1/2}\le Ce^{\delta_\mu {\rm Re}\, z}.$$
\noindent2) For $t\in\mathbb R$, the limit $$\pmb{\langle} [J_i,J_j(t)]\pmb{\rangle}_\infty=\lim_{\beta\to\infty}\lim_{L\to\infty}L^{-2}\langle [J_i,J_j(t)]\rangle_{\beta,L}$$
exists, uniformly in $t\in\mathbb R$. This follows from the Lieb-Robinson bounds on the quantum evolution of lattice systems \cite{LR}, and in particular
from one of its corollary, worked out in \cite{BrP, NOS}, which implies the existence of the infinite volume, zero temperature, 
real time correlations for a large class of quantum lattice systems. 

\medskip

\begin{figure}[ht]
\centering
\includegraphics[width=0.6\textwidth]{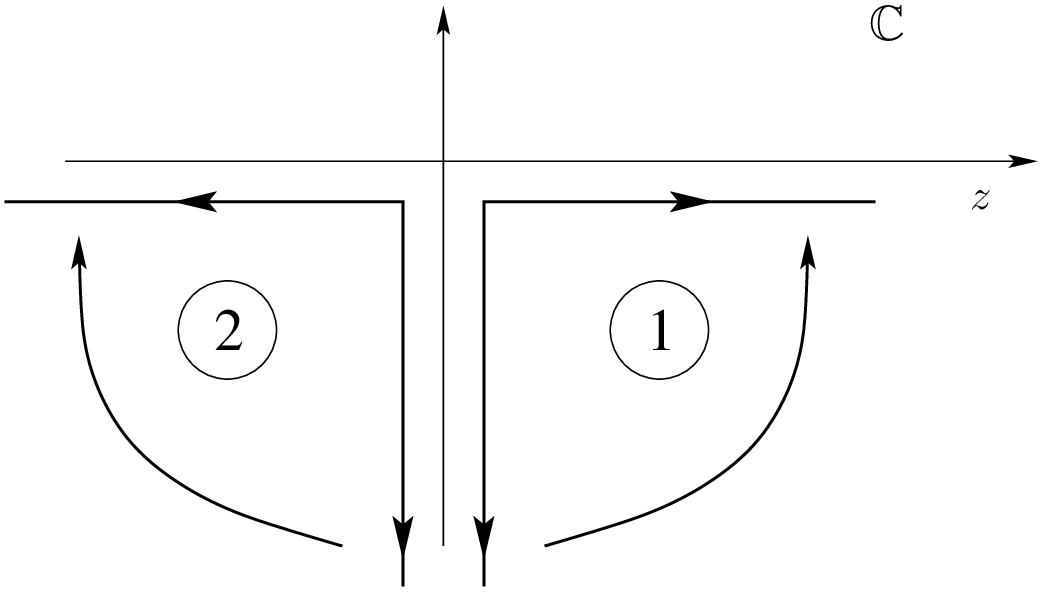}
\caption{\it The complex rotations of the integration paths performed on the two contributions in the right side of \eqref{kijo}. The rotation labelled (1)
(resp. (2)) refers to the first (resp. second) term in the right side of the equation.}
\label{fig.7}
\end{figure}

Given these two results, and using the fact that $\langle J_i\rangle_{\beta,L}=0$, one rewrites, for $\omega<0$,
\begin{eqnarray}
      \hat K_{ij}(\omega) &=& \lim_{\beta\to\infty}\lim_{L\to\infty}\frac1{L^2}\hfill\label{kijo}\>\times\\
&&      \int_0^{\beta/2}\Big(e^{-i\omega t}\langle
      J_{i}(-it)J_j\rangle_{\beta,L}+e^{+i\omega t}\langle
      J_{j}(-it)J_i\rangle_{\beta,L}\Big)dt \nonumber
\end{eqnarray}
Since the integrand, thought of as a function of $z=-it$, is analytic in the complex lower half-plane, and since it decays exponentially in $-{\rm Re}\,z$, one can rotate the integrals over the negative
imaginary axis as shown in {\it Fig. \ref{fig.7}}.

After the rotation shown in the picture, one rewrites: 
$$
\displaylines{
\hat K_{ij}(\omega) = \lim_{\epsilon\to
  0^+}\lim_{\beta\to\infty}\lim_{L\to\infty}\frac{i}{L^2} \> \times
\hfill\cr\hfill
\int_{-\beta/2}^0\Big(e^{-\omega(t+i\epsilon)}\langle J_{i}(-t-i\epsilon)J_j\rangle_{\beta,L}-
e^{-\omega (t-i\epsilon)}\langle
J_{j}(t-i\epsilon)J_i\rangle_{\beta,L}\Big)dt\
}$$
The two terms under the integral sign represent complex time current- current correlations, with times having a small imaginary part: in other words, the integration paths shadow the real axis from below. 
In order to `push' the paths on the real axis one uses the information that the real-time correlation exist, thanks to the Lieb-Robinson bounds. Using this fact, together with the a priori bounds on the complex
time correlations discussed above, one finds that the limit as $\epsilon\to0^+$ of the above integrals exist, and are equal to the time integral of the limiting (real-time) correlation functions, as desired. For details, see \cite[Section 6]{GMP17}. 

In conclusion, as announced in Section \ref{sec.5.1}, the first term in the right side of the Kubo conductivity \eqref{eq.0} equals $-\frac1\eta\hat K_{ij}(-\eta)$. Moreover, as 
claimed in the same section, the diamagnetic term equals $\frac1\eta\hat K_{ij}(0)$ (for a proof of this fact, based on the use of Ward Identities, see \eqref{diam} below), thus leading to the desired identity
$\sigma_{ij}=\bar\sigma_{ij}$, with $\bar\sigma_{ij}$ as in 
 \eqref{barsigma}.

\section{Ward Identities, Schwinger-Dyson equations and cancellations in perturbation theory}

From the theory of the Euclidean correlations discussion above, we inferred that $\sigma_{ij}=\bar \sigma_{ij}$, that 
$\bar\sigma_{ij}$ is analytic in $U$ in a small ball centered at the origin, and that the truncated multi-point Euclidean current correlations decay exponentially to zero at large space-time separation. I now want to combine these ingredients in order to show that $\bar \sigma_{ij}=\bar \sigma_{ij}\big|_{U=0}$, that is, 
all the interaction corrections to the Euclidean Kubo conductivity vanishes, which implies Theorem 1.

As anticipated above, these remarkable cancellations can be proved via a combination of two classes of identities, namely the Ward Identities and the Schwinger-Dyson equations, which are briefly reviewed in the following.

\subsection{Ward Identities}\label{sec7.1}

The Ward Identities that I want to discuss here are a consequence of the continuity equation for the lattice current. The starting point is the definition of the imaginary-time evolution of the fermionic density:
$n^\sigma_{(t,x)}:=e^{tH}n_{x,\sigma} e^{-tH}$. By deriving it with respect to $t$ one gets: 
\begin{equation}\label{cons.s}\frac{\partial}{\partial t}n^\sigma_{(t,x)}=\big[H,n^\sigma_{(t,x)}\big]=i\sum_{y,\sigma'}\Big(J^{\sigma\sigma'}_{xy}\Big)_t,
\end{equation}
where $\Big(J^{\sigma\sigma'}_{xy}\Big)_t$ is the imaginary time evolution of the lattice current: 
$$J^{\sigma\sigma'}_{xy}:=i\big(\psi^+_{x,\sigma}H^0_{\sigma,\sigma'}(x-y)\psi^-_{y,\sigma'}-\psi^+_{y,\sigma'}H^0_{\sigma',\sigma}(y-x)\psi^-_{x,\sigma}\big).$$
Note that by letting $\sigma$ vary in $I$ in \eqref{cons.s}, one obtains $|I|$ independent local conservation laws.
It is convenient to consider the following linear combination of these $|I|$ conservation laws: let $\tilde J_{0,(t,p)}:=q\sum_{x,\sigma}e^{-ipx_\sigma}n^\sigma_{(t,x)}$, where $x_\sigma=x+r_\sigma$; 
by deriving with respect to $t$, and using the anti-symmetry of the lattice current under the exchange $(x,\sigma)\leftrightarrow (y,\sigma')$, one finds: 
$$\frac{\partial}{\partial t}\tilde J_{0,(t,p)}=\frac{iq}{2}\sum_{\substack{x,y\\ \sigma,\sigma'}}(e^{-ipx_\sigma}-e^{-ipy_{\sigma'}})\Big(J^{\sigma\sigma'}_{xy}\Big)_t\equiv -p\cdot \tilde J_{(t,p)},$$
where in the last equation I defined $\tilde J_{(t,p)}$ to be
$$\tilde J_{(t,p)}:=\frac{q}2\sum_{\substack{x,y\\ \sigma,\sigma'}}e^{-ipx_\sigma}\eta\big(p(y_{\sigma'}-x_\sigma)\big)\,%^{\sigma\sigma'}_{xy}(p)
(y_{\sigma'}-x_\sigma)\Big(J^{\sigma\sigma'}_{xy}\Big)_t,$$
with $\eta(x):=(1-e^{-ix})/(ix)$. After Fourier transform with respect to (w.r.t.) the imaginary time variable, and letting ${\bf p}:=(\omega,p)$, this conservation law reads
\begin{equation}i\omega \hat J_{0,{\bf p}}+p\cdot \hat J_{\bf p}=0,\label{eq.cons}\end{equation}
where $\hat J_{0,{\bf p}}$ is the Fourier transform w.r.t. $t$ of $\tilde J_{0,(t,p)}$, and similarly for $\hat J_{\bf p}$, whose components will be denoted $\hat J_{i,{\bf p}}$, $i=1,2$.

\medskip

${\bf Remark}$.  The variable $p$ is introduced here as an auxiliary variable: the equations I will eventually be interested in 
are obtained by deriving w.r.t. $p$ the Ward Identities for the correlations of $\hat J_{\bf p}$ with other observables, and then 
setting this momentum variable to zero. Note that $\tilde J_{(t,p)}\big|_{p=0}=J_t$, which clarifies the connection with the current observable introduced above. 

\medskip

The desired Ward Identities are obtained by plugging the conservation law \eqref{eq.cons} into the formula for the multi-point current correlations: let 
\medium
\begin{equation}\label{eq.corr_def}\hat K_{\alpha_1,\ldots,\alpha_n}({\bf p}_1,\ldots,{\bf p}_{n-1}):=\lim_{\beta\to\infty}\lim_{L\to\infty}\frac1{\beta L^2}\langle T\, \hat J_{\alpha_1,{\bf p}_1};\cdots;\hat J_{\alpha_1,{\bf p}_1}\rangle_{\beta,L},
\end{equation}
\normalsize
where: $\alpha_i\in\{0,1,2\}\cup I$ and, if $\alpha\in\{0,1,2\}$, $\hat J_{\alpha,{\bf p}}$ has already been defined above, while, if $\alpha=\sigma\in I$, then $\hat J_{\sigma,{\bf p}}\equiv  \hat n^\sigma_{\bf p}$,
that is the Fourier transform w.r.t. $t$ of $\big(n^\sigma_x\big)_t$. Moreover, in the right side of \eqref{eq.corr_def}, ${\bf p}_n:=-{\bf p}_1\cdots -{\bf p}_{n-1}$. Note that $\hat K_{i,j}\big((\omega,0)\big)$ equals the 
Euclidean current-current correlation $\hat K_{ij}(\omega)$ defined in \eqref{Komega} above.
By using \eqref{eq.cons} into the definition \eqref{eq.corr_def} one gets
\begin{eqnarray}
i\omega_1\hat K_{0,\alpha_2,\cdots,\alpha_n}({\bf p}_1,&\ldots&,{\bf
  p}_{n-1})
+\sum_{i=1}^2 p_{1,i}\hat K_{i,\alpha_2,\cdots,\alpha_n}({\bf p}_1,\ldots,{\bf p}_{n-1})
\nonumber\\
&&\quad
=\sum_{j=2}^n\hat S_{\alpha_j;\hat \alpha_j}({\bf p}_1,\ldots,{\bf
  p}_{n-1}),\label{WI}
\end{eqnarray}
where $\hat \alpha_j=(\alpha_2,\ldots,\not\!\alpha_j,\ldots,\alpha_n)$ and, if I let 
\begin{equation}\hat \Delta_{\alpha_j}({\bf p}_1,{\bf p}_j):=\int_0^\beta dt\, e^{-i(\omega_1+\omega_j)t}\big[\tilde J_{0,(t,p_1)},\tilde J_{\alpha_j,(t,p_j)}\big],\label{Delta}\end{equation}
then 
\begin{equation}
  \vcenter{\displaylines{
      \hat S_{\alpha_j;\hat\alpha_j}({\bf
        p}_1,\ldots,{\bf p}_{n-1}):=
      \lim_{\beta,L\to\infty}\frac1{\beta L^2} \>\times
\hfill\cr\hfil
 \langle
      T\,\hat \Delta_{\alpha_j}({\bf p}_1,{\bf p}_j);\hat
      J_{\alpha_2,{\bf p}_2};\cdots; \hat J_{\alpha_{j-1},{\bf
          p}_{j-1}};\Big/\!\!\!\!\hat J_{\alpha_{j},{\bf p}_{j}};
      \cdots;\hat J_{\alpha_n,{\bf p}_n}\rangle_{\beta,L}.  \cr }}
\label{ESSE}
\end{equation}
In order to prove \eqref{WI}, one can start from the very definition of $i\omega_1\hat K_{0\alpha_2\cdots\alpha_n}$: 
$$
\displaylines{
  i\omega_1\hat K_{0,\alpha_2,\cdots,\alpha_n}({\bf p}_1,\ldots,{\bf p}_{n-1})
  \hfill\cr\hfill
  =\lim_{\beta\to\infty}\lim_{L\to\infty}\frac1{\beta L^2}\int_0^\beta dt_1\, e^{-i\omega_1 t_1}\frac{\partial}{\partial t_1}
\langle T\, \tilde J_{0,(t_1,p_1)}; \hat J_{\alpha_2,{\bf p}_2};\cdots;\hat J_{\alpha_n,{\bf
    p}_n}\rangle_{\beta,L}.
}
$$
In the right side, the derivative w.r.t. $t_1$, when acting on
$$\langle T\, \tilde J_{0,(t_1,p_1)}; \hat J_{\alpha_2,{\bf
    p}_2};\cdots;\hat J_{\alpha_n,{\bf p}_n}\rangle_{\beta,L},$$ can
either act on  $\tilde J_{0,(t_1,p_1)}$, in which case we get
$$
-p_1\cdot \langle T\, \tilde J_{(t_1,p_1)}; \hat J_{\alpha_2,{\bf
    p}_2};\cdots;\hat J_{\alpha_n,{\bf p}_n}\rangle_{\beta,L},
$$
or on the theta functions that implement the action of the
time-orderding operator $T$, in which case we get
$$
\sum_{j=2}^n\langle T\, \big[\tilde J_{0,(t_1,p_1)}, \tilde J_{\alpha_j,(t_1,p_j)}\big]; \hat J_{\alpha_2,{\bf p}_2};\cdots;
\Big/\!\!\!\! J_{\alpha_j,{\bf p}_j};\cdots;\hat J_{\alpha_n,{\bf
    p}_n}\rangle_{\beta,L}.
$$
Putting together the two contributions, we get \eqref{WI}.

\medskip

Now, recalling the fact that the truncated Euclidean correlation functions of local operators decay exponentially in space and (Euclidean) time, 
one has that their Fourier transforms are analytic in ${\bf p}_i$ in a neighborhood of the origin. In particular, I can differentiate \eqref{WI} w.r.t. $p_1$ and then set $p_1=0$. 
Let me do so in a few special cases, which will be useful in the following. Consider \eqref{WI} in the case $n\ge 2$ and $\alpha_j\in I$, $\forall j\ge 2$, 
in which case it reduces to 
$$i\omega_1\hat K_{0,\underline\sigma}({\bf p}_1,\ldots,{\bf p}_{n-1})+\sum_{j=1}^2p_{1,j}\hat K_{j,\underline\sigma}({\bf p}_1,\ldots,{\bf p}_{n-1})=0,$$
where $\underline\sigma=(\alpha_2,\ldots,\alpha_n)$. The reason why the right side of the Ward Identities is zero is that $\tilde J_{0,(t_1,p_1)}$ commutes with $\tilde J_{\sigma_j,(t_1,p_j)}$, 
$\forall \sigma_j\in I$. By differentiating the previous identity w.r.t. $p_1$, and then setting $p_1=0$, one gets: 
\begin{eqnarray}
&&  \hat K_{j,\underline\sigma}\big((\omega,0),{\bf p}_2,\ldots,{\bf
    p}_{n-1}\big)\label{wi1}\\
&&\qquad =-i\omega \frac{\partial}{\partial p_{1,j}}\hat K_{0,\underline\sigma}\big((\omega,0),{\bf p}_2,\ldots,{\bf p}_{n-1}\big),\quad j=1,2.	\nonumber
\end{eqnarray}
Similarly, if $n\ge 3$, $\alpha_2=j'\in\{1,2\}$, and $(\alpha_3,\ldots,\alpha_n)\in I^{n-2}$, then the Ward Identity \eqref{WI} reduces to 
\begin{equation}\label{uff1}
  \vcenter{\openup1\jot\halign{
      \hfil$\displaystyle{#}$
      &$\displaystyle{#}$\hfil
      \cr
 i\omega_1\hat K_{0,j',\underline\sigma}({\bf p}_1,\ldots,{\bf p}_{n-1})+\sum_{j=1}^2p_{1,j}\hat
      K_{j,j',\underline\sigma}&({\bf p}_1,\ldots,{\bf p}_{n-1})=
      \cr
&\hat S_{j';\underline\sigma}({\bf p}_1,\ldots,{\bf p}_{n-1}),
      \cr
      }}
\end{equation}
while, if $\alpha_2=0$,
$$i\omega_1\hat K_{0,0,\underline\sigma}({\bf p}_1,\ldots,{\bf p}_{n-1})+\sum_{j=1}^2p_{1,j}\hat K_{j,0,\underline\sigma}({\bf p}_1,\ldots,{\bf p}_{n-1})=0,$$
or, by exchanging the roles of ${\bf p}_1$ and ${\bf p}_2$,
\begin{equation}\label{uff2}i\omega_2\hat K_{0,0,\underline\sigma}({\bf p}_1,\ldots,{\bf p}_{n-1})+\sum_{j=1}^2p_{2,j}\hat K_{0,j,\underline\sigma}({\bf p}_1,\ldots,{\bf p}_{n-1})=0.\end{equation}
Now, by differentiating \eqref{uff1} w.r.t. $p_1$ and then setting $p_1=0$, one gets
\begin{align}
  \hat K_{j,j',\underline\sigma}\big((\omega_1,0)&,{\bf
    p}_2,\ldots,{\bf p}_{n-1}\big)\cr
  &=-i\omega_1 
\frac{\partial}{\partial p_{1,j}}\hat K_{0,j',\underline\sigma}\big((\omega_1,0),{\bf p}_2,\ldots,{\bf p}_{n-1}\big)\cr
&\qquad+\frac{\partial}{\partial p_{1,j}}\hat S_{j';\underline\sigma}\big((\omega_1,0),{\bf p}_2,\ldots,{\bf p}_{n-1}\big),\label{oops}\end{align}
Moreover, by differentiating \eqref{uff2} w.r.t. $p_2$ and then setting $p_2=0$, one gets 
\begin{align}\label{ciaciao}\hat K_{0,j',\underline\sigma}&({\bf
    p}_1,(\omega_2,0),{\bf p}_3,\ldots,{\bf p}_{n-1})\cr
&  =-i\omega_2\frac{\partial}{\partial p_{2,j'}}\hat K_{0,0,\underline\sigma}
({\bf p}_1,(\omega_2,0),{\bf p}_3,\ldots,{\bf p}_{n-1}).
\end{align}
By plugging \eqref{ciaciao} into \eqref{oops}, we obtain
\begin{align}
\hat K_{j,j',\underline\sigma}&\big((\omega_1,0),(\omega_2,0),{\bf
  p}_3,\ldots,{\bf p}_{n-1}\big)\cr
&=-\omega_1\omega_2 
\frac{\partial^2}{\partial p_{1,j}\partial p_{2,j'}}\hat K_{0,0,\underline\sigma}\big((\omega_1,0),(\omega_2,0),{\bf p}_3,\ldots,{\bf p}_{n-1}\big)\cr
&\qquad+\frac{\partial}{\partial p_{1,j}}\hat S_{j';\underline\sigma}\big((\omega_1,0),(\omega_2,0),{\bf p}_3,\ldots,{\bf p}_{n-1}\big).\label{oops2}\end{align}
In the following, I intend to combine the identities \eqref{wi1} and \eqref{oops2} with the Schwinger-Dyson equation. 
Before I do so, let me conclude this section by proving that the diamagnetic term in the Kubo conductivity is equal to \hfill\break
$\frac1{\eta}\hat K_{ij}(0)$: consider \eqref{WI} with $n=2$ and $\alpha_2=j\in\{1,2\}$:
$$i\omega\hat K_{0,j}({\bf p})+p_{1}\hat K_{1,j}({\bf p})+p_{2}\hat K_{2,j}({\bf p})=\hat S_j({\bf p}), \qquad {\rm where}\qquad {\bf p}=(\omega,p).$$
Deriving this equation w.r.t. $p$ and then setting ${\bf p}=0$, one gets:
\begin{equation}\label{diam} \hat K_{i,j}({\bf 0})=\frac{\partial \hat S_j}{\partial p_i}({\bf 0}).\end{equation}
Recalling the definition of $\hat S_j({\bf p})$, eq.\eqref{ESSE}, I
can write
$$
\hat S_j({\bf p})=q\pmb{\langle} \big[\sum_{x,\sigma}e^{-ip\cdot
    x_\sigma}n^\sigma_{(0,x)},J_{j,-p}\big]\pmb{\rangle}_\infty
$$ 
(here I denoted by $J_{j,-p}$ the $j$-th component of $\tilde J_{(t,-p)}\big|_{t=0}$), so that 
$$
\displaylines{\qquad
\hat K_{i,j}({\bf 0})=-iq\pmb{\langle}
\big[X_i,J_{j}\big]\pmb{\rangle}_\infty
\hfill\cr\hfill
=q^2\pmb{\langle}
\big[X_i,[H,X_j]\big]\pmb{\rangle}_\infty=-q^2\pmb{\langle}
\big[[H,X_i],X_j\big]\pmb{\rangle}_\infty,
\qquad}$$
as desired.

\null\vskip 9pt
\subsection{Schwinger-Dyson equation}

The Schwinger-Dyson equation that I want to discuss is an interpolation identity relating the current-current correlation $\hat K_{i,j}$
of the interacting model, with interaction strength $U$, 
to higher point correlation functions (such as current-current-density-density correlations) at interaction strength $U'$, where $U'$ is an 
interpolation parameter between $0$ and $U$. 
More general Schwinger-Dyson equations can be obtained along the same lines outlined below, for different multi-point correlations. 

For the purpose of this and the following section, I add a superscript $U$ to the interacting current-current correlation at interaction strength $U$ 
(and similarly for the higher point correlations),
in order to emphasize its dependence upon this parameter. By the fundamental theorem of calculus, 
$$ \hat{K}^U_{i,j}({\bf p})= \hat{K}^0_{i,j}({\bf p}) + \int_{0}^{U} dU'\, \frac{d}{dU'} \hat{K}^{U'}_{i,j}({\bf p}). $$
Recall that $\hat{K}^{U'}_{i,j}$ is the current-current correlation computed in the Gibbs state with Hamiltonian 
$H=H_0-U'V-\mu N$, see \eqref{Ham}. Therefore, by computing the derivative in $U'$, one finds
\begin{equation}\label{interp} \hat{K}^U_{i,j}({\bf p})= \hat{K}^0_{i,j}({\bf p}) + \int_{0}^{U} dU'\, \hat{K}^{U'}_{i,j,V}({\bf p}.-{\bf p}), \end{equation}
where 
$$\hat K^{U'}_{i,j}({\bf p},-{\bf p})=\lim_{\beta\to\infty}\lim_{L\to\infty}\frac1{\beta L^2}\langle T\, \hat J_{i,{\bf p}};\hat J_{j,-{\bf p}};\int_0^\beta\!\! V_t\, dt \rangle^{U'}_{\beta,L},$$
and, again, the superscript $U'$ in the right side has been added to emphasize that the Gibbs measure is associated with the 
Hamiltonian $H=H_0-U'V-\mu N$, with interaction strength $U'$. 
 
Now, in the limit $\beta,L\to\infty$, $\int_0^\beta V_t\, dt$ formally tends to $$\int_{\mathbb R} \frac{dq_0}{2\pi}\int_{\mathcal B}\frac{dq}{|\mathcal B|} \sum_{\sigma,\sigma'\in I}\hat n^\sigma_{\bf q}\hat v_{\sigma,\sigma'}(q)\hat n^{\sigma'}_{-{\bf q}},$$ where
${\bf q}=(q_0,q)$; at finite $\beta$ and finite $L$, $\int_0^\beta V_t\, dt$ can be written in the same form, but 
the integrals over $q_0$ and $q$ should be interpreted as suitable, discrete, Riemann sum approximations. By using this explicit expression for 
$\int_0^\beta V_t\, dt$, one rewrites:
$$
\displaylines{
  \hat K^{U'}_{i,j,V}({\bf p})=\lim_{\beta\to\infty}\lim_{L\to\infty}\frac1{\beta L^2}\>\times
  \hfill\cr\hfill\times
  \int \frac{dq_0dq}{2\pi|\mathcal B|}\sum_{\sigma,\sigma'\in I} \hat v_{\sigma,\sigma'}(q)
\langle T\, \hat J_{i,{\bf p}};\hat J_{j,-{\bf p}};\hat n^\sigma_{\bf q}\hat n^{\sigma'}_{-{\bf q}}
 \rangle^{U'}_{\beta,L}.
}$$ 
where, again,  the integrals over $q_0$ and $q$ should be interpreted as appropriate Riemann sums.
Note that there is no semicolon between $\hat n^\sigma_{\bf q}$ and $\hat n^{\sigma'}_{-{\bf q}}$, {\it i.e.}, the correlation function in the right side is not `completely truncated'. Of course, if desired, 
one can rewrite it as a linear combination of (products of) completely truncated correlations. If I do so, and then perform the limit $\beta,L\to\infty$, I 
find that \eqref{interp} can be rewritten as:
\begin{eqnarray}\hat K^{U}_{i,j}({\bf p})&=&\hat{K}^0_{i,j}({\bf p})+ \int_{0}^{U} dU'\sum_{\sigma,\sigma'\in I}\times\label{SD}\\
&\times& \Biggl[
\int_{\mathbb R} \frac{dq_0}{2\pi}\int_{\mathcal B}\frac{dq}{|\mathcal B|} \hat v_{\sigma,\sigma'}(q)
\hat K_{i,j,\sigma,\sigma'}^{U'}({\bf p},-{\bf p},{\bf q})\nonumber\\
&+&2\hat v_{\sigma,\sigma'}(0)\hat K^{U'}_{i,j,\sigma}({\bf p},-{\bf p})\hat K^{U'}_{\sigma'}\nonumber\\
&+&2\hat v_{\sigma,\sigma'}(-p)\hat K^{U'}_{i,\sigma}({\bf p})\hat K^{U'}_{j,\sigma'}(-{\bf p})\Big)\Biggr]\nonumber.\end{eqnarray}

\vskip -9pt

\subsection{Cancellations and proof of the main theorem}

By combining the Ward Identity, or, better, the two corollaries of the Ward Identity, eqs.\eqref{wi1} and \eqref{oops2}, with the Schwinger-Dyson equation, one can easily prove the desired cancellations, 
leading to the claim of our main theorem. Let me denote by $\bar\sigma_{ij}^U$ the Euclidean Kubo conductivity at interaction strength $U$: 
\begin{equation}\label{uffuff}\bar\sigma^{U}_{ij}=\frac1{|\ell_1\wedge\ell_2|}\lim_{\omega\to0}\frac{\partial \hat K_{i,j}^{U}(\omega,0)}{\partial \omega}.\end{equation}
Remember that $\hat K_{i,j}^{U}({\bf p})$ is smooth (analytic!) in its argument. 
By using the Schwinger-Dyson equation \eqref{SD} in the right side of \eqref{uffuff}, one gets: 
\begin{eqnarray}
&\hskip-2em  \bar\sigma^{U}_{ij}&=\bar\sigma^{0}_{ij}+ \frac1{|\ell_1\wedge\ell_2|}\sum_{\sigma,\sigma'\in I}\int_{0}^{U} dU'
\times\label{SD2}\\
  &&\times\lim_{\omega\to 0}\,
  \Biggl\{\int_{\mathbb R} \frac{dq_0}{2\pi}\int_{\mathcal B}\frac{dq}{|\mathcal B|}\hat v_{\sigma,\sigma'}(q)\frac{\partial}{\partial\omega}
\hat K_{i,j,\sigma,\sigma'}^{U'}\big((\omega,0),-(\omega,0),{\bf q})\nonumber\\
&&\ + 2\hat v_{\sigma,\sigma'}(0)\frac{\partial}{\partial\omega}
\hat K^{U'}_{i,j,\sigma}\big((\omega,0),-(\omega,0)\big)\hat K^{U'}_{\sigma'}\nonumber\\
&&\ + 2\hat v_{\sigma,\sigma'}(0)\frac{\partial}{\partial\omega}\Big[\hat K^{U'}_{i,\sigma}\big((\omega,0)\big)\hat K^{U'}_{j,\sigma'}(-(\omega,0))
\Big]\Biggr\}.\nonumber\end{eqnarray}
I now focus separately on the three contributions displayed in the three lines of this equation. Let me start with the first term, {\it i.e.}, the one 
involving $$\frac{\partial}{\partial\omega}\hat K_{i,j,\sigma,\sigma'}^{U'}\big((\omega,0),-(\omega,0),{\bf q}\big).$$ By using \eqref{oops2}, one can rewrite 
\begin{eqnarray}\label{orr}
&&  \frac{\partial}{\partial\omega}\hat
  K_{i,j,\sigma,\sigma'}^{U'}\big((\omega,0),-(\omega,0),{\bf q})=
\\
&&  \qquad =-\frac{\partial}{\partial\omega}\Big[\omega^2\frac{\partial^2}{\partial p_{1,i}\partial p_{2,j}}
\hat K_{0,0,\sigma,\sigma'}^{U'}\big((\omega,0),-(\omega,0),{\bf q})\Big],\nonumber \end{eqnarray}
where I used the fact that $\frac{\partial}{\partial p_{1,i}}\hat S_{j;\underline\sigma}\big((\omega,0),-(\omega,0),{\bf q}\big)$ is independent of $\omega$, by the very definition of 
$\hat S_{j;\underline\sigma}$, cf. with \eqref{Delta} and \eqref{ESSE} and recall that in this case $\omega_1=\omega=-\omega_2$. 
By recalling the smoothness and boundedness properties of $\hat K_{0,0,\sigma,\sigma'}^{U'}$ and of its derivatives, 
we see that \eqref{orr} implies that $$\frac{\partial}{\partial\omega}\hat K_{i,j,\sigma,\sigma'}^{U'}\big((\omega,0),-(\omega,0),{\bf q})$$ tends to zero as $\omega\to0$. Correspondingly, the contribution to $\bar\sigma_{ij}^{U}$ from the second line of \eqref{SD2} vanishes (the only issue to be discussed is the exchange of the limit with the integral over $q_0$ -- for details, see \cite[Section 4]{GMP17}). 

A similar discussion shows that the contributions to $\bar\sigma^{U}_{ij}$ from the third and fourth lines of \eqref{SD2} are zero, as well, and 
this completes the proof of Theorem 1.

\section{Conclusions}

I reviewed a recent proof of universality of the Kubo conductivity for a class of interacting lattice fermions in two dimensions. `Universality' refers here to the independence of the 
transport coefficient from the strength of the electron-electron interaction: {\it i.e.}, the interacting conductivity is equal to the non-interacting one, provided the non-interacting Hamiltonian 
is chosen to have a spectral gap, and the strength of the interaction is small, compared to the gap size. The class of models which this result applies to includes an interacting version of the Haldane model. The 
nice feature of this model is that it displays a transition from a `topologically trivial' insulating phase (characterized by vanishing transverse Kubo conducitivity) to a non trivial one, 
as the parameters $\phi$ and $M$ characterizing the hopping term are varied. The main theorem of these notes, Theorem 1, proves that the same transition takes place in the presence of interactions, thus providing an example of `topological phase transition' in a model of interacting electrons. 

Building upon the ideas and methods reviewed here, in particular combining them with an infrared multiscale analysis of the 
ground state of the interacting fermionic system, 
one can even characterize the nature of transition line from the topologically non-trivial phase to the trivial one, and it turns out that, 
in the case of the interacting Haldane model, the critical theory belongs to the same universality class as the non-interacting one. Nevertheless, the interactions have the remarkable effect 
of renormalizing (`dressing') the transition line: repulsive interactions tend to enlarge the topologically non-trivial region. For details, see \cite{GJMP, GMP19}.
Constructive fermionic renormalization group methods have also been used to prove universality of the longitudinal (`optical') conductivity 
for interacting graphene \cite{GMP11} and, more in general, for the interacting Haldane model on the
renormalized critical line \cite{GJMP, GMP19}. 

A lot remains to be done, but it is likely that extensions of the methods reviewed here, and further exchange of ideas between the mathematical physics and condensed matter communities, 
will allow us to attack and solve new problems that are currently beyond the state of the art, most notably the universality of quantum transport coefficients in interacting electron systems, in the presence of edges
and disorder (in the case of systems with edges and no disorder, see \cite{AMP,MP} for recent progress). I hope that the next decades will also witness
advances in other challenging open problems in mathematical physics, such as the theory of the 
superconducting phase in interacting Fermi systems, of the condensed phase in interacting Bose systems, and of the ferromagnetic phase in ferromagnetic quantum spin systems.

\appendix
\section{The non-interacting Hall conductivity as the first Chern number of the Bloch bundle}\label{app.A}

In this appendix, I give a sketch of the proof of \eqref{eq.2}. I assume for simplicity that $r_\sigma=0$, $\forall\sigma\in I$, and I set $q=-1$. In this case, after Fourier transform,  one gets
$$J=\sum_{\sigma,\sigma'}\int_{\mathcal B}\frac{dk}{|\mathcal B|}\hat \psi^+_{k,\sigma}\partial_k\hat H^0_{\sigma,\sigma'}(k)\hat \psi^-_{k,\sigma'},$$
where $\hat \psi^\pm_{k,\sigma}=\sum_x e^{\mp ik\cdot x}\psi^\pm_{x,\sigma}$. Using the Wick rule, one gets 
$$
\displaylines{
\pmb{\langle} [J_i,J_j(t)]\pmb{\rangle}_\infty
 =\sum_{\sigma_1,\ldots,\sigma_4}\int_{\mathcal B}\frac{dk}{|\mathcal B|}\partial_{k_i}\hat H^0_{\sigma_1,\sigma_2}(k)
\>\times
 \hfill\cr\hfill
\pmb{\langle} \hat
\psi^-_{k,\sigma_2}\hat\psi^+_{k,\sigma_3}(t)\pmb{\rangle}_\infty
\partial_{k_j}\hat H^0_{\sigma_3,\sigma_4}(k)\pmb{\langle} \hat
\psi^+_{k,\sigma_1}\hat\psi^-_{k,\sigma_4}(t)\pmb{\rangle}_\infty
-\Big(i\leftrightarrow j\Big),
}$$
where
\begin{eqnarray}&&\pmb{\langle} \hat \psi^-_{k,\sigma_2}\hat\psi^+_{k,\sigma_3}(t)\pmb{\rangle}_\infty=\big[P_+(k)e^{i(\hat H^0(k)-\mu)t}\big]_{\sigma_2,\sigma_3},\\
&&
\pmb{\langle} \hat \psi^+_{k,\sigma_1}\hat\psi^-_{k,\sigma_4}(t)\pmb{\rangle}_\infty=\big[P_-(k)e^{-i(\hat H^0(k)-\mu)t}\big]_{\sigma_4,\sigma_1}.\label{A.2}\end{eqnarray}
I plug these expressions in the previous formula, and I use the fact that $$\partial_{k}\hat H^0(k)=\partial_k\sum_{\alpha=\pm}P_\alpha(k)\hat H^0(k)P_\alpha(k).$$ 
After a straightforward computation, one finds that $$\pmb{\langle} [J_i,J_j(t)]\pmb{\rangle}_\infty=G_{ij}''(t),$$
with 
$$\displaylines{
  G_{ij}(t)=
\hfill\cr\hfill
  -\int_{\mathcal B}\frac{dk}{|\mathcal B|}{\rm
  Tr}\Big[\partial_{k_i}P_-(k)e^{i(\hat
    H^0(k)-\mu)t}\partial_{k_j}P_-(k)\, P_-(k)e^{-i(\hat
      H^0(k)-\mu)t}\Big]
\hfill\cr\hfill\hfill
  +\Big(i\leftrightarrow j\Big).
\hfill}$$
The contribution to the Kubo conductivity involving $\pmb{\langle} [J_i,J_j(t)]\pmb{\rangle}_\infty$ can then be written as $\frac1{|\ell_1\wedge\ell_2|}$ times the limit as $\eta\to 0^+$ of 
$$\frac{-i}{\eta}\int_{-\infty}^0\, e^{\eta t} G_{ij}''(t) dt=\frac{-i}\eta G_{ij}'(0)+iG_{ij}(0)-i\eta\int_{-\infty}^0 e^{\eta t}G_{ij}(t)dt.$$
Now, the third term in the right side goes to zero as $\eta\to 0^+$.  Moreover, a straightforward computation shows that the first term in the right side (which is singular as $\eta\to 0^+$) is equal to 
$$\frac{-i}\eta G_{ij}'(0)=\frac{1}\eta \int_{\mathcal B}\frac{dk}{|\mathcal B|}{\rm Tr}\ \partial_{k_i}\hat H^0(k)\partial_{k_j}P_-(k),$$
which cancels exactly with the diamagnetic contribution to the Hall conductivity. In fact, after performing a Fourier transform, and using \eqref{A.2}, one finds
\begin{eqnarray}-\frac1{\eta}
\pmb{\langle} [[H,X_i],X_j]\pmb{\rangle}_\infty&=&\frac1{\eta}\int_{\mathcal B}\frac{dk}{|\mathcal B|}
\pmb{\langle} \hat \psi^+_{k}\partial_{k_i}\partial_{k_j}\hat H^0(k)\hat\psi^-_{k}\pmb{\rangle}_\infty\\
&=&\frac1{\eta}\int_{\mathcal B}\frac{dk}{|\mathcal B|}{\rm Tr}\ P_-(k)\partial_{k_i}\partial_{k_j}\hat H^0(k)\\
&=&-\frac1{\eta}\int_{\mathcal B}\frac{dk}{|\mathcal B|}{\rm Tr}\ \partial_{k_i}\hat H^0(k)\partial_{k_j}P_-(k),\end{eqnarray}
that is  $\frac{-i}{\eta}G_{ij}'(0)-\frac1{\eta}\pmb{\langle} [[H,X_i],X_j]\pmb{\rangle}_\infty=0$. 

In conclusion, plugging all these relations in \eqref{eq.0} at $U=0$, and recalling the fact that $|\ell_1\wedge\ell_2|\,|\mathcal B|=(2\pi)^2$, one finds
$$\displaylines{\quad
  \sigma_{ij}\Big|_{U=0}=\frac{i}{|\ell_1\wedge\ell_2|}G_{ij}(0)
\hfill\cr\hfill
  =-i\int_{\mathcal B}\frac{dk}{(2\pi)^2}{\rm Tr}\Big\{P_-(k)\big[\partial_{k_i}P_-(k),\partial_{k_j}P_-(k)\big]\Big\},
  \quad}
$$
as desired.

\medskip

${\bf Acknowledgments.}$ This review is based on a presentation given at the Istituto Lombardo, Accademia di Scienze e Lettere, in Milano (Italy) on May 5, 2016,
as well as on the notes of a course given at the EMS-IAMP summer school in mathematical physics {\it Universality, Scaling Limits and Effective Theories},
held in Roma (Italy) on July 11-15, 2016. I gratefully thank the members of the Istituto Lombardo, Accademia di Scienze e Lettere,
and in particular Dario Bambusi and Antonio Giorgilli, for their invitation to present my work on the occasion of an `Adunanza' of the Istituto Lombardo.
The research leading to these results has received funding from the European Research
Council under the European Union's Horizon 2020 Programme, ERC Consolidator Grant UniCoSM (grant
agreement n. 724939).


\begin{thebibliography}{999999}

\bibitem{AGD} A. A. Abrikosov, L. P. Gorkov, I. E. Dzyaloshinski: {\it Methods of quantum field theory in statistical physics}, Prentice-Hall 1963.

\bibitem{AG} M. Aizenman, G. M. Graf: {\it Localization bounds for an electron gas}, J. Phys.A: Math.Gen. {\bf 31}, 6783 (1998).

\bibitem{AM} J. F. Allen, A. D. Misener: {\it Flow of liquid helium II}, Nature {\bf 141}, 75 (1938).

\bibitem{An} P. W. Anderson: {\it The Theory of Superconductivity in the High-$T_c$ Cuprates}, Princeton Series in Physics, Princeton Univ. Press 1997.

\bibitem{AEMWC} M. H. Anderson, J. R. Ensher, M. R. Matthews, C. E. Wieman, E. A. Cornell: {\it Observation of Bose-Einstein condensation in a dilute atomic vapor},
Science {\bf 269}, n. 5221, 198-201 (1995).

\bibitem{AMP}  G. Antinucci, V. Mastropietro, M. Porta: {\it  Universal edge transport in interacting Hall systems}, Comm. Math. Phys. {\bf 362}, 295-359 (2018). 

\bibitem{Av} J. E. Avron: {\it Colored Hofstadter butterflies}, in: Multiscale methods in quantum mechanics,  Trends Math. (Boston, MA: Birkhauser Boston, 2004), pp. 11-22

\bibitem{AS} J. E. Avron, R. Seiler: {\it Quantization of the Hall conductance for general, multiparticle Schr\"odinger
Hamiltonians}, Phys. Rev. Lett. {\bf 54}, 259-262 (1985).

\bibitem{ASS} J. E. Avron, R. Seiler, B. Simon: {\it Homotopy and quantization in condensed matter physics}, Phys. Rev.
Lett. {\bf 51}, 51 (1983).

\bibitem{BCS1} J. Bardeen, L. N. Cooper, and J. R. Schrieffer: {\it Microscopic theory of superconductivity}, Phys. Rev. {\bf 106}, 162 (1957).

\bibitem{BCS2} J. Bardeen, L. N. Cooper, and J. R. Schrieffer: {\it Theory of superconductivity}, Phys. Rev. {\bf 108}, 1175 (1957).

\bibitem{BF} G. A. Battle, P. Federbush: {\it A note on cluster expansions, tree graph identities, extra 1/N! factors!}, Lett. Math. Phys. {\bf 8}, 55-57  (1984).

\bibitem{BM} J. G. Bednorz, A. M\"uller: {\it Possible high $T_c$ superconducitivity in the Ba-La-Cu-O system}, Z. Phys. B Cond. Matt. {\bf 64}, 189-193 (1986).

\bibitem{BVS} J. Bellissard, A. van Els, H. Schulz-Baldes: {\it The non-commutative geometry of the quantum Hall
effect}, J. Math. Phys. {\bf 35}, 5373 (1994).

\bibitem{Bo} N. N. Bogoliubov: {\it On the theory of superfluidity}, Izv. Akad. Nauk USSR {\bf 11}, 77 (1947). Eng. Trans. J. Phys. (USSR) {\bf 11}, 23  (1947).

\bibitem{BSTH} C. C. Bradley, C. A. Sackett, J. J. Tollett, R. G. Hulet: {\it Evidence of Bose-Einstein condensation in an atomic gas with attractive interactions},
Phys. Rev. Lett. {\bf 75}, 1687 (1995).

\bibitem{BrP} J. B.  Bru, W. A. de S.Pedra: {\it Lieb-Robinson Bounds for Multi-Commutators and Applications to Response Theory}, Springer Briefs in Mathematical Physics {\bf 13}, Springer, 2016. 

\bibitem{BrP2} J. B.  Bru, W. A. de S.Pedra: {\it Universal bounds for large determinants from non-commutative H\"older inequalities in fermionic constructive quantum field theory}, 
Math. Mod. Meth. Appl. Sc. {\bf 12} (27) (2017).

\bibitem{Br} D. C. Brydges: {\it A short course on cluster expansions}, Ph\'enom\`enes critiques, syst\`emes al\'eatoires, th\'eories de jauge, Part I, II (Les Houches, 1984), 129-183, North-Holland, Amsterdam, 1986.

\bibitem{BrF} D. C. Brydges, P. Federbush: {\it A new form of the Mayer expansion in classical statistical mechanics}, J. Math. Phys. {\bf 19}, 2064-2067 (1978).

\bibitem{BK} D. Brydges, T. Kennedy: {\it Mayer expansions and the Hamilton-Jacobi equation}, J. Stat. Phys. {\bf 48}, 19-49 (1987).

\bibitem{CH} S. Coleman, B. Hill: {\it No more corrections to the topological mass term in QED3}, Phys. Lett. B. {\bf 159}, 184 (1985).

\bibitem{CGS} M. Correggi, A. Giuliani, R. Seiringer:  {\it Validity
of spin wave theory for the quantum Heisenberg model}, Europhys. Lett. {\bf 108}, 20003 (2014); and {\it Validity of the spin-wave approximation for the free
energy of the Heisenberg ferromagnet}, Comm. Math. Phys. {\bf 339}, 279-307 (2015).

\bibitem{DMA} K. B. Davis, M.-O. Mewes, M. R. Andrews, N. J. van Druten, D. S. Durfee, D. M. Kurn, W. Ketterle: {\it Bose-Einstein condensation in a gas of sodium atoms},
Phys. Rev. Lett. {\bf 75}, 3969 (1995).

\bibitem{GeM} G. Gentile, V. Mastropietro: {\it Renormalization group for one-dimensional fermions. A review on mathematical results}, Phys. Rep. {\bf 352} 273-437 (2001).

\bibitem{GJMP}  A. Giuliani, I. Jauslin, V. Mastropietro, M. Porta: {\it Topological phase transitions and universality in the Haldane-Hubbard model}, Phys. Rev. B {\bf 94}, 205139 (2016).

\bibitem{GM} A. Giuliani, V. Mastropietro: {\it The two-dimensional Hubbard model on the honeycomb lattice}, Comm. Math. Phys. {\bf 293}, 301-346 (2010). 

\bibitem{GMP11} A. Giuliani, V. Mastropietro, M. Porta: {\it Absence of interaction corrections in the optical
conductivity of graphene}, Phys. Rev. B {\bf 83}, 195401 (2011); and {\it Universality of conductivity in interacting graphene},
Comm. Math. Phys. {\bf 311}, 317-355 (2012).

\bibitem{GMP17} A. Giuliani, V. Mastropietro, M. Porta: {\it Universality of the Hall Conductivity in Interacting
Electron Systems}, Comm. Math. Phys. {\bf 111}, 1107-1161 (2017).

\bibitem{GMP19} A. Giuliani, V. Mastropietro, M. Porta: {\it Quantization of the interacting Hall conductivity in the critical regime}, J. Stat. Phys. (2019). https://doi.org/10.1007/s10955-019-02405-1. 

\bibitem{Ue} C. Golschmidt, D. Ueltschi, P. Windridge: {\it Quantum Heisenberg models and their probabilistic representations},
Contemp. Math. {\bf 552}, Entropy and the quantum, II, R. Sims and D. Ueltschi, Ed. 2011.

\bibitem{Hal} F. D. M. Haldane: {\it Model for a quantum Hall effect without Landau levels: condensed-matter
realization of the `Parity Anomaly'}, Phys. Rev. Lett. {\bf  61}, 2015 (1988).

\bibitem{Hall1879} E. Hall: {\it On a New Action of the Magnet on Electric Currents}, American Jour. Math. {\bf 2} (3), 287-292  (1879).

\bibitem{Ha} B. I. Halperin: {\it Quantized Hall conductance, current-carrying edge states, and the existence of extended states in a two-dimensional disordered potential},
Phys. Rev. B {\bf 25}, 2185 (1982).

\bibitem{Has} M. B. Hastings: {\it The stability of free Fermi Hamiltonians},
J. Math. Phys. {\bf 60}, 042201 (2019).

\bibitem{HM} M. B. Hastings and S. Michalakis: {\it Quantization of Hall Conductance for Interacting Electrons
on a Torus}, Comm. Math. Phys. {\bf 334}, 433-471 (2015).

\bibitem{HJV} I. F. Herbut, V. Juricic, and O. Vafek: {\it Coulomb Interaction, Ripples, and the Minimal Conductivity of Graphene}, Phys. Rev. Lett. {\bf 100}, 046403
(2008); V. Juricic, O. Vafek, and I. F. Herbut: {\it Conductivity of interacting massless Dirac particles in graphene: Collisionless regime}, Phys. Rev. B {\bf 82},
235402 (2010); I. F. Herbut, V. Juricic, and O. Vafek: {\it Comment on `Minimal conductivity in graphene: Interaction corrections and ultraviolet anomaly' by Mishchenko E. G}, e-print
arXiv:0809.0725.

\bibitem{Ho} D. R. Hofstadter: {\it Energy levels and wave functions of Bloch electrons in rational and irrational
magnetic fields}, Phys. Rev. B {\bf 14}, 2239-2249  (1976).

\bibitem{Ka} P. Kapitza: {\it Viscosity of liquid helium below the $\lambda$-point}, Nature {\bf 141}, 74 (1938).

\bibitem{Kl} K. von Klitzing, G. Dorda, M. Pepper: {\it New method for high-accuracy determination of the fine-structure constant based on quantized Hall resistance}, Phys. Rev. Lett {\bf 45}, 494 (1980).

\bibitem{La} R. B. Laughlin: {\it Quantized Hall conductivity in two dimensions}, Phys. Rev. B {\bf 23}, 5632(R) (1981).

\bibitem{La2} R. B. Laughlin: {\it Anomalous Quantum Hall Effect: An Incompressible Quantum Fluid with Fractionally Charged Excitations}, Phys. Rev. Lett. {\bf 50}, 1395 (1983).

\bibitem{LR} E. H. Lieb, D. W. Robinson: {\it The finite group velocity of quantum spin systems}, Comm. Math.
Phys. {\bf 28}, 251-257 (1972).

\bibitem{LSSY} E. H. Lieb, R. Seiringer, J.-P. Solovej, J. Yngvason: {\it The Mathematics of the Bose Gas and Its
Condensation}, Oberwolfach Seminars {\bf 34}, Birkhauser, Basel (2005).

\bibitem{MP} V. Mastropietro, M. Porta: {\it  Spin Hall insulators beyond the Helical Luttinger model}, Phys. Rev. B {\bf 96}, 245135 (2017).

\bibitem{NOS} B. Nachtergaele, Y. Ogata, R. Sims: {\it Propagation of correlations in quantum lattice systems}, J. Stat. Phys. {\bf 124}, 1-13 (2006); B. Nachtergaele, R. Sims: {\it Lieb-Robinson bounds in quantum many-body physics}, Contemp. Math. {\bf 529}, 141-176 (2010).

\bibitem{NO} J. W. Negele, H. Orland: {\it Quantum many-particle systems}, Addison-Wesley Pub. Co., 1988.

\bibitem{On} H. K. Onnes: {\it The resistance of pure mercury at helium temperatures}, Commun. Phys. Lab. Univ. Leiden {\bf 12}, 120 (1911).

\bibitem{PS08} W. A. de S. Pedra, M. Salmhofer: {\it Determinant Bounds and the Matsubara UV Problem of Many-Fermion Systems}, Comm. Math. Phys. {\bf 282}, 797-818 (2008).

\bibitem{STG} H. L. Stormer, D. C. Tsui, A. C. Gossard: {\it The fractional quantum Hall effect}, Rev. Mod. Phys. {\bf 71}, S298 (1999).

\bibitem{Th} D. J. Thouless: {\it Localisation and the two-dimensional quantum Hall effect}, J. Phys. C {\bf 14}, 3475 (1981).

\bibitem{TKNdN} D. J. Thouless, M. Kohmoto, M. P. Nightingale, and M. den Nijs: {\it Quantized Hall Conductance in a Two-Dimensional Periodic Potential}, Phys. Rev. Lett. {\bf 49}, 405
(1982).

\bibitem{TDG} D. C. Tsui, H. L. Stormer, A. C. Gossard: {\it Two-dimensional magnetotransport in the extreme quantum limit}, Phys. Rev. Lett {\bf 48}, 1559 (1982).
 
\end{thebibliography}
\end{document}